\documentclass[a4paper,10pt]{article}
\pdfoutput=1
\usepackage[english]{babel}

\usepackage{float}

\usepackage{tikz}
\usepackage{slashed}
\usepackage{relsize}
\usepackage{MnSymbol} 
\usepackage{authblk}
\usepackage{lineno}   

\usepackage{siunitx}
\usepackage{booktabs}

\usepackage{zref-xr,zref-user}
\zexternaldocument*[SM-]{zz}

\usepackage[hypertexnames=true]{hyperref}

\newcommand*{\MM}[1]{\mathsmaller{\mathsmaller{#1}}}

\newcommand*\circled[1]{\tikz[baseline=-2.5pt]{  %baseline=(char.base)
            \node[shape=circle,draw,minimum width=.25cm,minimum height=.25cm, inner sep=0pt] (char) {#1};}}

\newcommand*{\sloop}[1]{\circled{$\MM{#1}$}}        

\DeclareRobustCommand{\sloopcaption}[1]{\circled{$\MM{#1}$}} 

\newcommand*{\twocycle}[1]{\mathop{ {\mathsmaller{\upfilledspoon}}\mkern-15mu{\mathsmaller{\downfilledspoon}}  }\limits_{#1}}
\newcommand*{\twocyclec}[2]{\mathop{ {\mathsmaller{\textcolor{#1}{\upfilledspoon}}}\mkern-15mu{\mathsmaller{\textcolor{#1}{\downfilledspoon}}}  }\limits_{#2}}

\newcommand*{\pntwocyclec}[2]{\mathop{ {\textcolor{#1}\downfootline}\mkern-15mu{\textcolor{#1}\uparrow}  }\limits_{#2}}
\newcommand*{\pptwocyclec}[2]{\mathop{ {\textcolor{#1}\downarrow}\mkern-15mu{\textcolor{#1}\uparrow}  }\limits_{#2}}

\newcommand*{\diff}[1]{\mkern-4mu\mathop{ \mathlarger{\mathlarger{{\uprsquigarrow}}}}\limits_{#1}\mkern-4mu}

\newcommand*{\diffc}[2]{\mkern-4mu\mathop{\mathlarger{\mathlarger{{\textcolor{#1}{\uprsquigarrow}}}}}\limits_{#2}\mkern-4mu}
\newcommand*{\qdiffc}[3]{q^{#3}\mkern-4mu\mathop{\mathlarger{\mathlarger{{\textcolor{#1}{\uprsquigarrow}}}}}\limits_{#2}}
\newcommand*{\difft}[1]{\mkern-4mu\mathop{ \mathlarger{\mathlarger{{\uprsquigarrow}}}}_{#1}\mkern-0mu}

\newcommand*{\xsp}[1]{\mkern #1mu }
\newcommand*{\pxsp}[1]{\mkern #1mu+\mkern #1mu }
\newcommand*{\mxsp}[1]{\mkern #1mu-\mkern #1mu }

\title{\textbf{Key features of Turing systems are determined purely by network topology}}

\author[1,2]{Xavier Diego}
\author[3]{Luciano Marcon}
\author[3]{Patrick M{\"u}ller}
\author[1,2,4]{James Sharpe}

\affil[1]{\small EMBL-CRG Systems Biology Research Unit, Center for Genomic Regulation, Barcelona Institute for Science and Technology, Barcelona, Spain}
\affil[2]{\small Universitat Pompeu Fabra, Barcelona, Spain}
\affil[3]{ \small Friedrich Miescher Laboratory of the Max Planck Society, Germany}
\affil[4]{ \small Institucio Catalana de Recerca i Estudis Avancats, Barcelona, Spain}

\date{\vspace{-5ex}}

\begin{document}  %**************
\maketitle        %**************
%\linenumbers

\begin{abstract}
Turing's theory of pattern formation is a universal model for self-organization, applicable to many
systems in physics, chemistry and biology. Essential properties of a Turing system, such as the conditions for 
the existence of patterns and the mechanisms of pattern selection are well understood in small networks. 
However, a general set of rules governing how network topology determines fundamental system  
properties and constraints has not be found. Here we provide a first general theory of Turing network topology,
which proves why three key features of a Turing system are directly determined by the topology: 
the type of restrictions that apply to the diffusion rates, the robustness of the system, and the phase relations of the molecular species.
\end{abstract}

\section{Introduction}
The hallmark of biological development is the formation of spatially organized cellular structures. 
In 1952, Alan Turing proposed a mechanism based on the reaction and diffusion of morphogen molecules that would allow 
cells to self-organize and form periodic patterns \cite{Turing1952}. However, nearly 40 years passed until Turing patterns
were observed in the CIMA chemical reaction \cite{Castets1990,Ouyang1991}. 
The main reason why Turing patterns had been so elusive was that they can not occur if all 
the molecules diffuse at the same rates \cite{Murray2003}, as typically occurs in laboratory reactions.
Further, Turing models with moderate diffusion ratios require a level of adjustment in the
reaction parameters that is unrealistic \cite{Baker2008}, an issue that has been refereed to as the fine-tuning 
problem \cite{Butler2011}. The severity of these requirements has cast doubts about the relevance of Turing patterns in biological systems. 
In the CIMA reaction, the diffusion constraint was serendipitously circumvented by the introduction 
of an immobile color indicator that reversibly bound to one of the reactants and slowed down its diffusion \cite{Lengyel1991, Agladze1992}. 
Hindering the diffusion of the activator with a non-diffusible complexing agent was the basis of a method proposed
to systematically design new Turing reactions \cite{Lengyel1992}. A refinement of this method \cite{Horvath2009} has
been followed in the design of almost all new chemical systems producing Turing patterns \cite{Szalai2015}
and inspired most of the theoretical efforts to relax the diffusion constraints
in Turing networks \cite{Pearson1992a, Korvasova2015}. Turing patterns with equal diffusion rates of the diffusible molecules have also 
been noticed in models of biological pattern formation that include immobile cell membrane receptors
as part of the network \cite{Rauch2004, Levine2005, Strier2007, Klika2012}. 
These networks have the common feature of relying on the introduction of a non-diffusible node that interacts with the activator but is
inert to the other reactants, following the architecture of the original CIMA model. 

\begin{figure} [!h]
%\captionsetup[subfigure]{labelformat=simple, labelsep=parens}%{labelformat=empty}
\centering
{\includegraphics[width=0.7\textwidth]{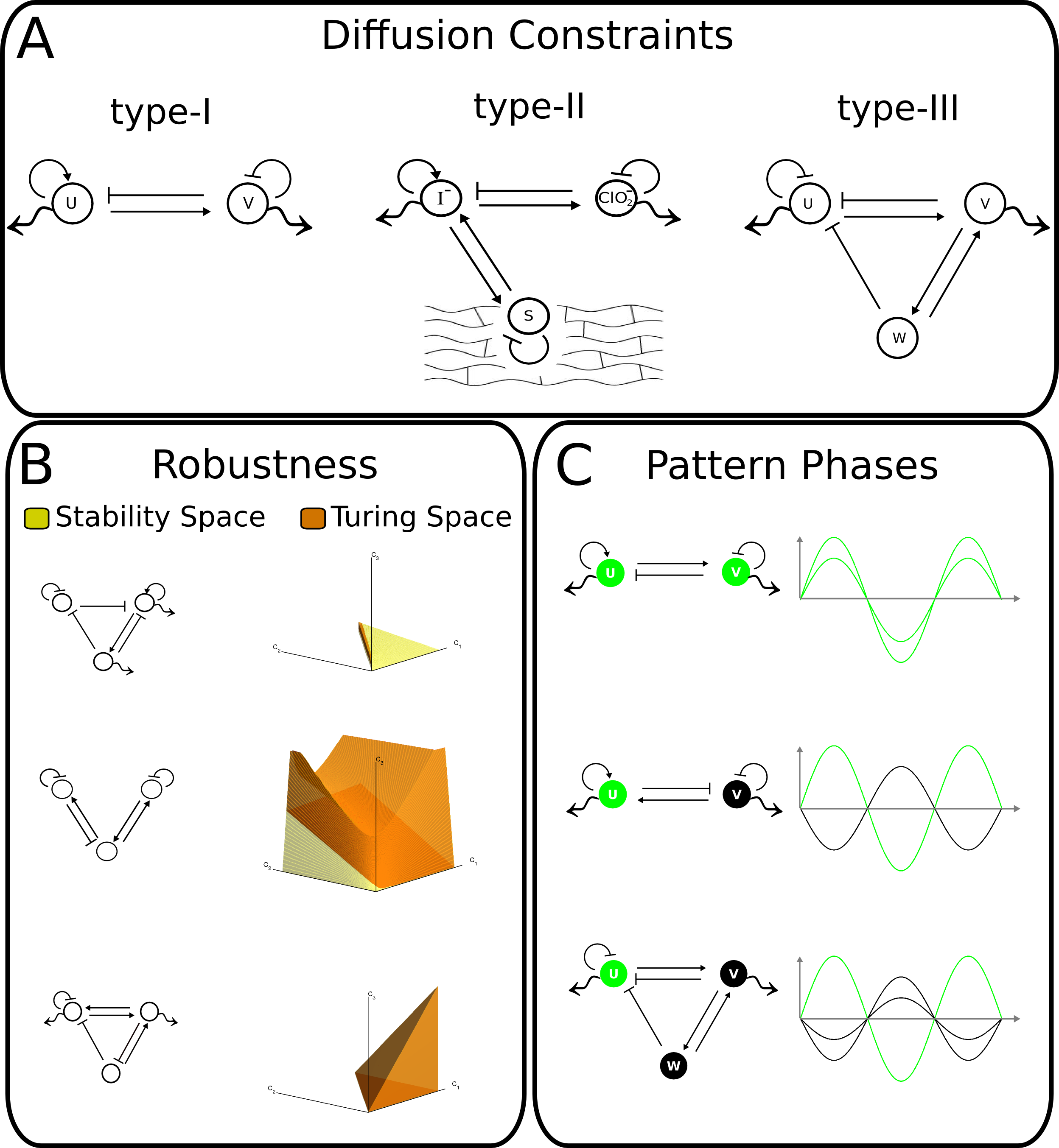}}
\caption{Three questions under investigation a) Why the diffusion rates of $u$ and $v$ in the standard 2-node Turing network (left) 
must be very different, whereas in the CIMA network (center) iodine and chlorite can diffuse at the same rate and in the third (right) network 
the diffusion rates of $u$ and $v$ are completely unconstrained? b) Why the robustness of these three Turing networks is so different? c) 
What determines the phase of each species in a Turing pattern?}
\label{fig1}
\end{figure}

However, in a recent study \cite{Marcon2016} we have found that the hindered diffusion architecture 
is just one particular case amongst many other possibilities for the relaxation of diffusion constraints 
in Turing networks with immobile nodes. 
Our computational analysis of all possible reaction-diffusion networks of 3 and 4
nodes revealed that they can be classified into three types according to the restrictions that apply to the diffusion rates. 
The first type comprises networks in which a subset of the species must diffuse at a higher rate than the 
rest, as is the case in the classical 2-node Turing networks. In the second type, the diffusion rates 
are subjected to certain constraints but can form Turing patterns even if the diffusion rates of the mobile species are all equal, 
as in the CIMA reaction \cite{Lengyel1991} and related models \cite{Korvasova2015, Rauch2004, Levine2005}.
The third type is formed by networks in which the diffusion rates of the mobile species are not subjected to any constraint,
a novel class of Turing networks that had not been found before. 
These computational results suggested that central aspects of Turing networks have to be clarified.\\
Here we demonstrate that the type of diffusion constraints that apply to Turing 
network of any number of nodes are determined by its topology. 
Topology also explains a new class of pseudo-patterning networks that we call Turing filters.
The patterns generated by Turing filters do not have a characteristic wave-length; instead these 
networks amplify preexisting spatial heterogeneities if their characteristic wavelength is smaller than a critical threshold.
Also, the graph analysis allows us to distinguish 
networks that can undergo oscillatory Turing instabilities, and the classification according to diffusion 
constraints carries over to these patterning systems. 
Secondly, the analysis shows that 
Turing networks can be grouped into a few topological families, and that the robustness associated to the size of their Turing space is largely
determined by them. 
Finally, our analysis allows us to resolve a question that, surprisingly, 
has not been addressed before: what determines the phase overlaps of the species in a Turing pattern? 
Again, the graph structure of a network allows us to predict the phases of the species and it shows also
how to construct a network with any desired combination of phase overlaps.

 \subsection*{Methods: Graph theory for Turing networks}
      A network of interacting species whose concentration changes through 
      local reactions and spatial diffusion can be described by a set of reaction-diffusion equations. %p1: The problem to solve

      \begin{equation}
      \label{eq:RDeqs}
      \frac{{\partial r_i }}
      {{\partial t}} = f_i ( {r}) + {d_i }
      \nabla ^2 r_i \,\,\,\,\,\,\,\,\,\,\,\,\,i = 1,...,N
      \end{equation}
      where $r_i(x,t)$, $f_i$ and $d_i\geqslant 0$ represent the concentrations, reaction 
      rates and non-negative diffusion constants. The system is assumed to be stable without diffusion 
      and there is no flow of reactants outside a finite domain. Generally, diffusion smooths out spatial heterogeneities in these type of systems.  
      Turing's genius intuition \cite{Turing1952} consisted in realizing that diffusion could have
      the opposite effect if the reactants interacted in the appropriate way, so
      that a spatially periodic pattern would replace the homogeneous state as the stable equilibrium.
      Existence of Turing patterns is demonstrated by analyzing the evolution of a system under small perturbations,
      which can be predicted from the linear approximation of the reaction-diffusion equations \cite{Casten1977}. 
      This leads to an eigenvalue problem that reduces the derivation of the conditions for diffusion-driven instability
      to the analysis of the zeroes of the characteristic polynomial $P_q(\lambda)=\det[\lambda{\bf{I}}-({\bf{J}^R}(r_o)-q^2{\bf{D}})]$, 
      where ${\bf{J}^R}(r_0)$ is the Jacobian of the reaction term evaluated at equilibrium and $\bf{D}$ is the diffusion matrix.
      The departure from equilibrium is a superposition of periodic modes of wavenumber $q$ and speed of growth
      or decay given by the eigenvalues $\lambda(q)$. If the real part of the largest $\lambda(q)$ is positive, this mode grows exponentially and 
      contributes to the emergence of a Turing pattern. The mode $q_c$ associated to the eigenvalue with the maximum real part grows faster and dominates the final pattern.
      If $\lambda(q_c)$ is real, a stationary pattern with wavelength $2\pi/q_c$ emerges. 
      If $\lambda(q_c)$ is complex, an oscillatory pattern emerges with wavelength $2\pi/q_c$ 
      and oscillation period given by $2\pi/Im(\lambda(q_c))$. The eigenvalues are given by the zeroes of $P_q(\lambda)$. 
      For a network with $N$ species $P_q(\lambda)$ is a polynomial of degree $N$ in $\lambda$:
      \begin{equation}
      \label{eq:Pq}
      P_{q} (\lambda ) = \lambda ^N  + a_1(q) \lambda ^{N - 1}  + ... + a_{N - 1} (q)\lambda  + a_N(q)=0  
      \end{equation}                     
      where the coefficients $a_i(q)$ are functions of the kinetic constants and the diffusion rates.
      The location of the zeroes of a polynomial in the complex plane
      is given by the Routh-Hurwitz theorem, but simpler conditions to locate them can be derived in terms 
      of the coefficients  $a_k(q)$ \cite{Gantmacher1959}. Stability without diffusion requires that all the coefficients of $P_q(\lambda)$
      are positive for $q = 0$. Hence $a_K (0) > 0$ for $K = 1, ..., N$ is a necessary condition for stability and, conversely, $a_K (q)<0$ for some $K$ is a sufficient condition for diffusion-driven instabilities.      
      In turn, necessary conditions
      for the existence of stationary Turing patterns can be derived in terms of the sign of $a_N(q)$: 
       
      \begin{equation}
      \label{eq:conditionsS}
      \begin{array}{*{20}c}
      {\exists \,q > 0,\,\,\forall\,k < N} & | & \begin{gathered}
	 a_k (q) > 0  \,  \\
	\,a_N (q) < 0\,  \\ 
      \end{gathered}  &  \Leftarrow  & {stationary\, Turing}  \\
      \end{array} 
      \end{equation} 
 
 For oscillatory patterns, a similar condition can be derived in terms of  $a_{K<N}(q)$, but it is only sufficient \cite{Clarke1993,Mincheva2007}:
      
      \begin{equation}
      \label{eq:conditionsO}
      \begin{array}{*{20}c}
      {\exists \,q > 0,\,\,\exists\,k < N} & | & \begin{gathered}
	a_k (q) < 0  \,  \\
	\,a_N (q) > 0\,  \\ 
      \end{gathered}  &  \Rightarrow  & {oscillatory\,Turing}  \\
      \end{array} 
      \end{equation}

      A comprehensive discussion of the derivation and scope of the conditions for diffusion-driven instabilities is given in SM1. In principle, the parametric constraints for the existence of Turing patterns 
      can be derived analytically from these conditions. In practice, they become intractable for networks with
      more than 3 diffusible species. 
       
       \begin{figure}    [H]
      \centering
      {\includegraphics[width=0.7\linewidth]{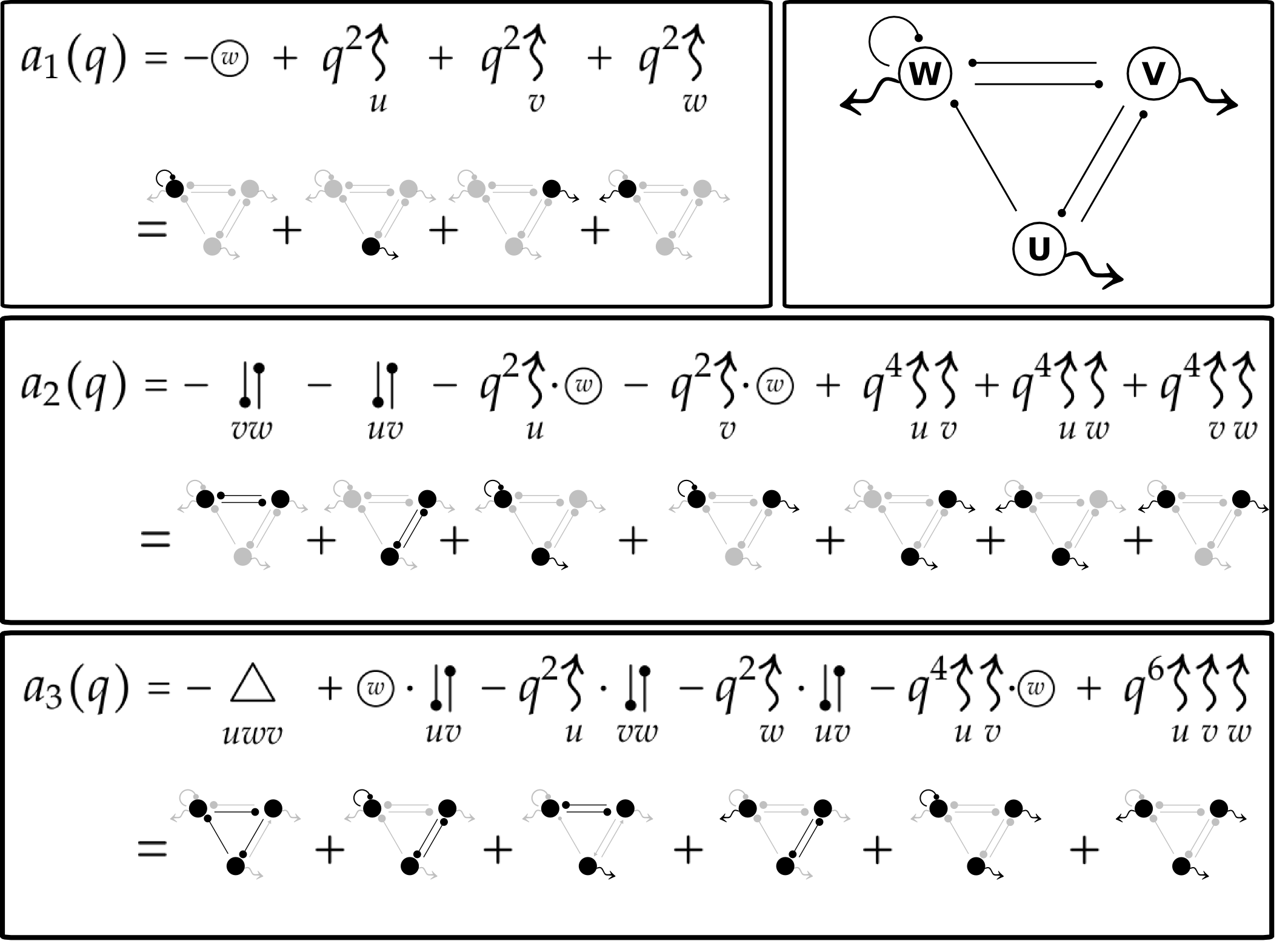}}
      \caption{Calculation of $P_q(\lambda)$ as a sum of $\ell$-subgraphs. Nodes, edges and diffusive loops that form each $\ell$-subgraph
      are shown in black. The loop of the $i$-th node is represented by \sloopcaption{i} in the equations.}  % Background of $\ell_{RD}$-subgraphs involving diffusion  loops is shaded
      \label{fig:fig3}
      \end{figure} 
         For this reason, we introduce a method based on Graph theory to recast them in terms of the topology of the underlying reaction-diffusion system. 
      In this way we reveal a connection between the structure of a reaction-diffusion system and diffusion constraints, robustness and pattern phases.
       To that end, a directed graph is associated to the matrix ${\bf{F}^{RD}}(q)={\bf{J}}^R(r_0)-q^2{\bf{D}}$ obtained from the
      linear approximation of the reaction-diffusion equations. Our definition of the reaction-diffusion graph 
      follows the definition of the Coates graph of a square matrix: the graph of a
      reaction-diffusion network with $N$ species has $N$ nodes and an
      edge from the $j$-th node to the $i$-th node if the entry ${\bf{J}^R}(r_o)_{ij}$ is non-zero \cite{Brualdi08}. 
      The entries on the diagonal of the Jacobian result in edges that start and end in the same node. These edges
      are called loops and are associated to decay or self-activation terms. 
      In addition, for each non-zero entry in $\bf{D}$, a special type of loop represented by a wriggled arrow is added to the
      corresponding diffusible node. The weight of each edge is given by the corresponding entry in ${\bf{F}^{RD}}(q)$. Particularly, the diffusive
      loop associated to the diffusible node $i$ has weight $-q^2 d_{i}$. The fundamental elements of the graph are cycles. A cycle of length
      $m$ is a set of $m$ edges that form a closed path joining $m$ distinct nodes. By this definition, loops are cycles of length one. The
      weight of a cycle is defined as the product of its edges. 
      Cycles are classified as positive or negative according to the sign of their weight. Two important graph structures are Induced subgraphs
      and Linear spanning subgraphs, or for short, $\ell$-subgraphs. The Induced subgraph of $k$ nodes is formed by these nodes and all
      the edges between them. Conversely, the complementary nodes of an induced subgraph of size $k$ are the $N-k$ nodes that are not contained in it.
      An $\ell$-subgraph of size $k$ is a set of disjoint cycles that spans $k$ nodes and is contained in their
      Induced subgraph. 
      The weight of an $\ell$-subgraph is: 
      \begin{equation}
      \label{eq:wl}
      w(\ell ) =  \prod\limits_{c \subseteq \,\,\ell } {(-w(c))  } 
      \end{equation}
      Thus, the weight of an $\ell$-subgraph is positive if it is formed by negative cycles or contains an even number of positive cycles.
      In this case it is said to be an stabilizing $\ell$-subgraph.
      Examples of cycles, induced and $\ell$-subgraphs and the association of the reaction-diffusion graph for a 4-node network are shown in SM2. 
      Importantly, $\ell$-subgraphs are the only contributors to $P_q(\lambda)$,
      as it can be proven from the Laplace expansion of the characteristic polynomial \cite{Horn1990} and the Coates expression for the
      determinant \cite{Brualdi08}. Precisely, the coefficient $a_k(q)$ is given by the
      sum of all the $\ell$-subgraphs of size $k$ in the reaction-diffusion graph. The expression of the coefficients of
      $P_q(\lambda)$ in terms of subgraphs for a minimal 3-node network is shown \ref{fig:fig3}. The contribution of each Induced subgraph of $k$ nodes
      can be separated into a) $\ell_R$-subgraphs formed only by reaction cycles b) mixed $\ell_{RD}$-subgraphs formed by $m$ diffusive loops
      and the complementary $\ell_R$-subgraph formed by reaction
      cycles spanning the other $k-m$ nodes. c) an $\ell_D$-subgraph formed by $k$ diffusive loops, provided that all the inducing nodes are diffusible:
  
      \begin{equation}
      \label{eq:a_k}
       \begin{split}
      a_k (q) =&\sum\limits_{I_{\gamma _k}} \bigg[ \sum\limits_{\ell _R \subseteq I_{\gamma _k} } {w(\ell_R )}  
	      \,\,+\sum\limits_{\gamma _m  \subset \,\gamma _k }^{m < k}  {q^{2m} \Big( \prod\limits_{j \in \gamma _m } {d_j }\Big) }  \cdot 
		\Big( \sum\limits_{\ell _R  \subseteq I_{\gamma _k  - \gamma _m } } w(\ell_R )  \Big)\\
	      &+ q^{2k} \prod\limits_{j \in \gamma _k } {d_j }  \bigg] 
      \end{split}
      \end{equation}
      
      Two important results follow from the previous expression. 
      First, the topology of a network, understood as the distribution of cycles, cycle signs and diffusion loops, 
      determines exclusively the requirements for the existence of Turing patterns. 
      The reason is that the conditions for diffusion-driven instability depend
      on the coefficients $a_k(q)$ and these are functions of $\ell$-subgraphs only.
      Therefore, the existence of Turing patterns imposes constraints on the relative weights of cycles, 
      rather than individual kinetic parameters. 
      Second, a Turing network must have a destabilizing module: an Induced subgraph in which the destabilizing
      $\ell_R$-subgraphs outweigh the stabilizing $\ell_R$-subgraphs. %Destabilizing$\ell$-subgraphs have an odd number of positive cycles and stabilizing $\ell$-subgraphs have none or an even number of them.
      Typically, this condition requires that there is a set of nodes linked by a positive cycle that
      outweighs any stabilizing $\ell$-subgraphs contained in their Induced subgraph. If a network does not have destabilizing module, all 
      the terms in $a_N(q)$ are positive and the condition \ref{eq:conditionsS} for instability can not be fulfilled. 
      A rigorous proof of this result is given in SM3 and constitutes a generalization of the requirement of a self-activator in 2-node Turing networks.

      \section{Results}
\subsection{Topology and the source Diffusion constraints}
In our previous work \cite{Marcon2016} we developed a symbolic algebra procedure to 
      obtain the exhaustive list of 3 and 4-node networks with the minimal number of edges that can generate 
      stationary Turing patterns. The analysis revealed that 3-node and 4-node networks with two diffusible nodes can be classified into three types according to 
      the diffusion constraints for the generation of Turing patterns. Defining the ratio of diffusion rates of
      the two diffusible species as $d$, and $p$ as the space of kinetic parameters compatible with 
      Turing patterns, the constraints for each Type can be stated as:
      \begin{equation}
      \label{eq:types}
      \begin{array}{*{20}l}
	{\text{Type - I\,\,\,\,\,: }}   & {\forall p} & {d > 1}  \\
	{\text{Type - II\,\,\,:}}  & {\exists p} & {d = 1}  \\
	{\text{Type - III\,:}} & {\forall p} & {d > 0}  
      \end{array} 
      \end{equation}
      Surprisingly, we found that there are as many 3-node networks with one immobile reactant of Type-II and Type-III   
      as of Type-I, whereas the 4-node networks with two immobile reactants of Type-III outnumber the networks of Type-I and Type-II. 
      In other words, against the widely held belief, Turing networks with mild or no diffusion constraints are very common. 
      Here we demonstrate how the topology of a network explains these results. 
      According to the stability condition, all the independent terms $a_k(0)$ in eq.\ref{eq:a_k} 
      are positive. The leading terms in $a_k(q)$, if present, are also strictly positive for $q>0$.
      According to the condition for Turing instability,
      $a_N(q)$ must cross the zero and turn negative for some $q>0$. % Diffusion constraints in Turing networks stem from the opposite requirements of stability without diffusion and instability with diffusion. 
      Decartes’ rule of signs provides an upper bound for the 
      number of real positive zeros of a real polynomial \cite{Struik2014}. Particularly, a polynomial with only non-negative
      coefficients cannot have real positive zeros. It then follows that $a_N(q)$ must have a negative coefficient, 
      and the negative coefficient must lie at some intermediate degree in $q^2$. 
      This is a necessary condition for Turing instabilities.
      In a network in which all the species diffuse, this is only possible with differential diffusivity. 
      The algebraic proof of this well known result is given in SM9 but it does not 
      reveal the source of the requirement and how it can be weakened. To that end it is necessary
      to examine how the network graph leads to a nested structure of the characteristic polynomial.
      As shown in eq.\ref{eq:a_k}, the coefficient of degree $2m$ in $a_N(q)$ is the sum of all mixed $\ell_{RD}$-subgraphs formed
      by $m$ diffusive loops and an $\ell_R$-subgraph  
      that spans the other $N-m$ nodes of the network. Importantly, each of these $\ell_R$-subgraphs of size $N-m$ contribute also 
      to the coefficient $a_{N-m}(0)$ of $a_{N-m}(q)$.
      For example, in the topology shown in \ref{fig:fig3}
      the coefficient $a_3(q)$ is: 
      
      \begin{equation}
      \label{eq:a3q}
      a_3(q)= a_3(0)- \qdiffc{black}{u}{2}\cdot\twocycle{vw}\mxsp{2}
		      \qdiffc{black}{w}{2}\cdot\twocyclec{black}{uv}
		    -\qdiffc{black}{u}{4}\diffc{black}{v}\cdot\sloop{w}
		    \pxsp{2}\qdiffc{black}{u}{6}\diffc{black}{v}\xsp{3}\diffc{black}{w}
      \end{equation}
      
      Thus, the coefficient of degree $q^2$ in $a_3(q)$ contains all the reaction $\ell_R$-subgraphs of size 2  
      that form the independent term $a_2(0)$ in $a_2(q)$. Likewise, the coefficient of
      degree $q^4$ in $a_3(q)$ contains all the loops that form $a_1(0)$ in $a_1(q)$. This illustrates the nested structure of the $P_q(\lambda)$:
      
      \begin{equation}
      \label{eq:a120}
      \begin{split}
      a_1(0)&=-\sloop{w} \xsp{90} \\
      a_2(0)&=-\twocyclec{black}{vw}\mxsp{5}\twocycle{uv} 
      \end{split}
      \end{equation} 
      
      Stability imposes that $a_k(0)>0$ and therefore, that the stabilizing subgraphs outweigh the destabilizing subgraphs of size $k$. 
      It follows that for any of the coefficients
      of intermediate degree in  $a_N(q)$ to be negative, the diffusion loops complementary of the destabilizing 
      $\ell_R$-subgraphs have to compensate this difference.
      Thus, the differential diffusion requirement for Turing instabilities
      stems from the necessary condition derived from Decartes' rule of signs. 
      To illustrate the relationship explicitly, let the Induced subgraph of $u$ and $v$ be the destabilizing module in the network from \ref{fig:fig3}, and the cycle of 
      length two between them the only positive cycle. 
      Then, only the coefficient of degree $q^2$ in $a_3(q)$ can be negative. 
      Imposing this and assuming for simplicity that the diffusion rates of the nodes in the destabilizing module are equal, 
      the constraint on the diffusion ratio $d=\difft{w}/\difft{u}=\difft{w}/\difft{v}$ takes the following form:  
      
      \begin{equation}
      \label{eq:constraint1}
      \left. {\begin{array}{*{20}c}
	-\pntwocyclec{black}{vw}\mxsp{5}\pptwocyclec{red}{uv}\, > \,0  \\
	{}\\
	-\pntwocyclec{black}{vw}\cdot \diff{u}\mxsp{5}\pptwocyclec{red}{uv}\cdot \diff{w} \,< \, 0 \\
      \end{array}} \right\} \Rightarrow d\,\, > \xsp{2}-\pntwocyclec{black}{vw}/\pptwocyclec{red}{uv}\xsp{2}>\xsp{2}1
      \end{equation}
      
      Two observations about the diffusion constraints are in order. 
      First, the constrains on diffusion rates that stem from Decartes' rule are necessary for the existence of Turing patterns, but not sufficient. 
      If the necessary ratio is set, 
      the sufficient conditions are obtained imposing that $a _N(q)$ turns negative, 
      which results in additional requirements for the kinetic parameters but not for the diffusion rates. 
      Second, the nested structure of the characteristic polynomial and Decartes' rule imposes
      that at least one of the species complementary to the destabilizing module has a larger diffusion rate than the species that induce it,
      and never the other way around. This is the generalization of the
      requirement of differential diffusion between an activator and an inhibitor in $2$-node networks:
      in larger networks, the role of the activator and inhibitor cannot be assigned to individual species,
      but to network subgraphs.
      Importantly, the previous argument carries over for general networks of any size but depends on the assumption that all species diffuse. Each coefficient $a_k(0)$ is formed by all the reaction $\ell_R$-subgraphs 
      of size $k$. If all species diffuse, each $\ell_R$-subgraph in $a_k(0)$ can be 
      coupled to $m$ diffusive loops of complementary nodes to form a mixed
      $\ell_{RD}$-subgraph of size $k+m$ that contributes to the coefficient of degree $2m$ in $a_{k+m}(q)$. 
      Thus, the nested structure of the characteristic polynomial,
      from which the diffusion constrains stem, is a general property 
      of networks in which all species diffuse. Because of this, all networks in which
      all species diffuse belong to the Type-I class.
 \subsection*{Relaxation of diffusion constrains}
      In networks with immobile species, the nested structure of the characteristic polynomial does not necessarily hold. 
       \begin{figure} [H]
      \centering
      {\includegraphics[width=0.7\linewidth]{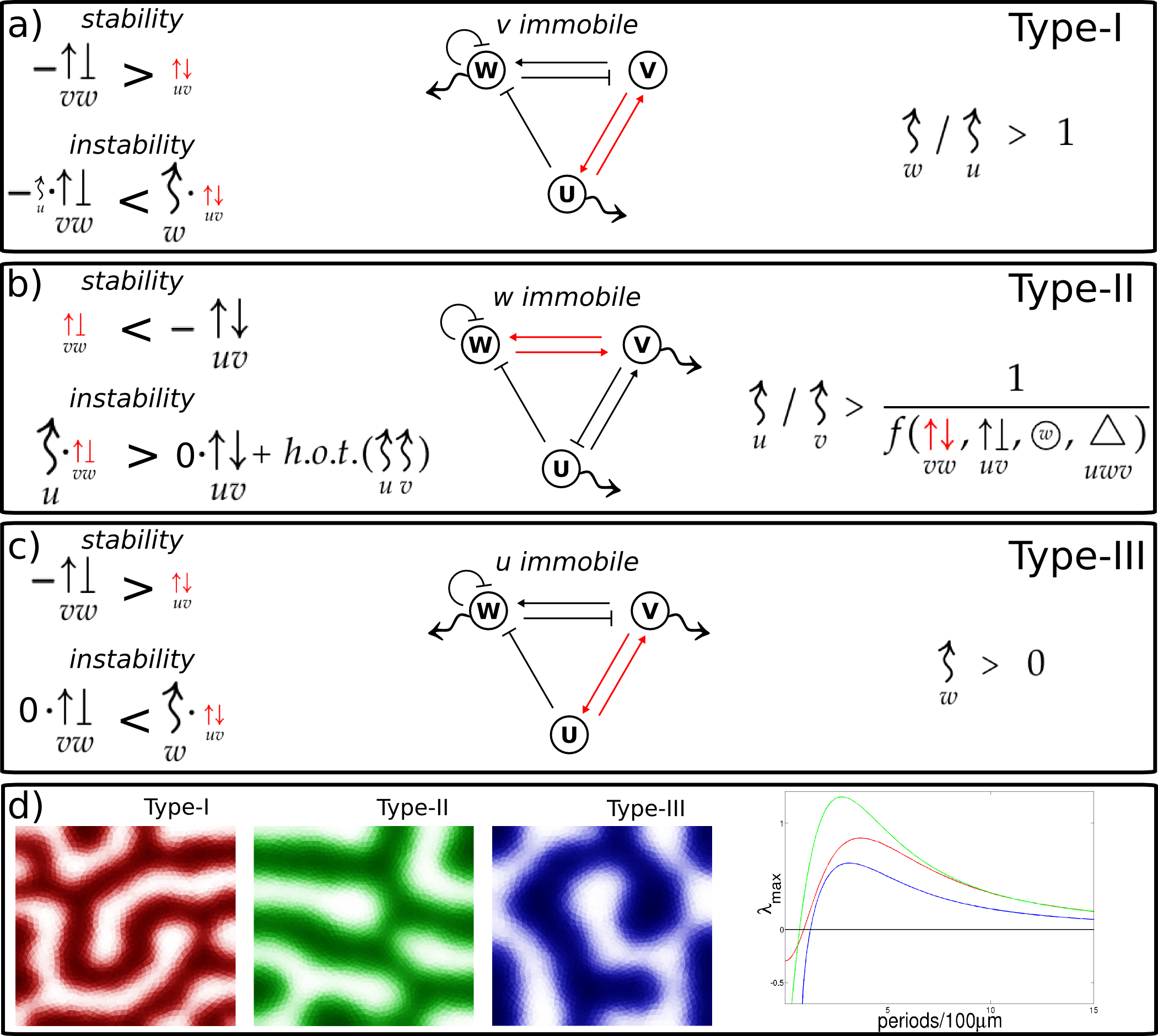}}
      \caption{ a-c) Permutation of the immobile node in Topology of \ref{fig:fig3}.
      results in a Turing network of each Type. Stability and instability requirements are shown on the left, diffusion constraints on the right.   
      d) Patterns and dispersion relationships 
      of Type-I, II, III are qualitatively similar (shown in red, green and blue) } 
      \label{fig:fig5}
      \end{figure}
       This property is lost if there is at least one $\ell_R$-subgraph with one or more complementary nodes that are non-diffusible.     
        Assuming that this subgraph is of size $k$, it will contribute to $a_k(0)$ but it will not to $a_N(q)$. 
      Particularly, it will be missing from the coefficient of degree $2m$ in $a_N(q)$ formed by products of $\ell_R$-subgraphs 
      of size $k$ and $m=N-k$ diffusion loops of their complementary nodes. If the subgraph that is missing is stabilizing and the destabilizing module is also of size $k$, the requirements on the diffusion rates stemming from Decartes’ rule are weakened. The destabilizing module can outweigh the remaining subgraphs and make the coefficient of degree $2m$ in $a_N(q)$ negative, even if the diffusion loops of its complementary nodes are equal or smaller than its own. In this way, the necessary condition of differential diffusivity is weakened.\\       
      By the same mechanism, subgraphs of smaller size than the destabilizing module might be prevented from contributing to coefficients of higher degree than $2m$ in $a_N(q)$.
      In turn, this facilitates the fulfillment of the sufficient condition for the existence of Turing patterns. 
      This is the principle that underlies the relaxation of diffusion constraints and the associated classification of Turing networks. The precise  
      topological characterization of each Turing Type is given in table \ref{tab:types}:
      
      \begin{table}[h]
\begin{tabular}{|c|c|}%\hline
\toprule
Type-I & \multicolumn{1}{m{0.85\linewidth}|}{All stabilizing $\ell_R$-subgraphs of the same size as the destabilizing module have all their complementary nodes diffusible} \\ \midrule
Type-II & \multicolumn{1}{m{0.85\linewidth}|}{At least one stabilizing $\ell_R$-subgraph of the same size as the destabilizing module has an immobile complementary node} \\ \midrule
Type-III &  \multicolumn{1}{m{0.85\linewidth}|}{The destabilizing module is the $\ell_R$-subgraph of smallest size that has all its complementary nodes diffusible}\\%\hline
\bottomrule
\end{tabular}
\caption{\label{tab:types} Topological features of Turing networks}
\end{table}

      The topological properties of the different Turing Types results in algebraic differences that allow 
      to make a simple distinction based on the form of the characteristic polynomial. 
      In Type-III networks the destabilizing module is the only contributor to the 
      leading term in $a_N(q)$. Hence, for all modes with wavenumber $q$ above a certain threshold $a_N(q)$ turns negative and the system is unstable 
      independently of the diffusion rates. 
      Both the necessary condition derived from Decartes' rule and the 
      sufficient condition $a_N(q)<0$ are guaranteed by the topology.\\
      There are two configurations that result in a Type-II network. In the first configuration, 
      the destabilizing module is the only contributor of its size to a coefficient in $a_N(q)$, but there are stabilizing subgraphs of smaller size that 
      contribute to the coefficients of larger degree. It follows that the necessary condition is guaranteed by the 
      topology but the sufficient condition still involves the diffusion rates. 
      In the second configuration of Type-II networks, the destabilizing module is not the only contributor to a coefficient in $a_N(q)$, but
      at least one stabilizing subgraph of the same size is missing. Thus, the necessary condition is not guaranteed by the topology, 
      but it can be fulfilled without differential diffusion. Whether there are terms of higher degree or 
      not determines if the sufficient condition is satisfied automatically or if it imposes additional requirements on the kinetic rates. 
      In Type-I networks, all subgraphs of the same size as the destabilizing module contribute to the corresponding term in $a_N(q)$. 
      Hence, the necessary condition imposes differential diffusion. Again, if there are coefficients of higher degree, they further restrict the space of parameters compatible with Turing instability.\\
      \indent To illustrate the principle for the relaxation of diffusion constraints, one node of a minimal topology at a time is 
      assumed to be immobile to obtain a Turing network of different Type according 
      to the diffusion constraints. The same procedure is applied in SM4 to $4$-node and non-minimal networks to demonstrate the power of the graph-based
      framework to analyze the relaxation of diffusion constraints in complex networks. 
      The CIMA reaction \cite{Lengyel1991,Lengyel1992} and the relaxation principle operating in several models from the literature \cite{Rauch2004,Levine2005,Klika2012,Korvasova2015} are also analyzed in SM5. 
      The topology shown \ref{fig:fig3} has all nodes diffusible and therefore can only produce Type-I Turing networks.
      If the subgraph induced by $u$ and $v$ is the destabilizing module and $v$ is assumed to be immobile, the coefficient $a_3(q)$ 
      given in eq.\ref{eq:a3q} is reduced to:
      \begin{equation}
      \label{eq:Type-I}
      a_3(q)= a_3(0) -\qdiffc{black}{u}{2}\cdot\pntwocyclec{black}{vw}
		      -\qdiffc{black}{w}{2}\cdot\pptwocyclec{red}{uv}  
      \end{equation}
      Hence, the network is still a Type-I Turing system limited by the constraint given in eq.\ref{eq:constraint1}: the diffusion rate of $w$ must be bigger than that of $u$, otherwise the coefficient of  
      degree $q^2$ cannot be negative. If this occurs, the sufficient condition $a_3(q)$ is fulfilled automatically for sufficiently large wavenumbers.
      Conversely the same topology  becomes a Type-II network assuming that 
      $w$ is immobile and that the subgraph induced by $w$ and $v$ is the destabilizing module.
      Then, the coefficient $a_3(q)$ is:
      \begin{equation}
      \label{eq:Type-II}
      a_3(q)= a_3(0) -\qdiffc{black}{u}{2}\cdot\pptwocyclec{red}{vw}
		      -\qdiffc{black}{u}{4}\diffc{black}{v}\cdot\sloop{w}           
      \end{equation}
      Thus, the destabilizing module is the only
      contributing term to the coefficient of degree $2$ in $a_3(q)$ and the necessary condition that stems from Decartes' rule is 
      fulfilled automatically and independently of the diffusion ratio $d$. Because there is a coefficient of larger degree
      in $q$, fulfillment of the sufficient condition $a_3(q)<0$ involves the diffusion ratio, but as expected from a Type-II network, 
      it can be satisfied with $d=1$.\\
      Finally, the topology from \ref{fig:fig3} is transformed into a Type-III network assuming that the subgraph induced by $u$ and $v$
      is the destabilizing module and that $u$ is not diffusible. In this case, the destabilizing module is the only contributor to the
      leading coefficient in $a_3(q)$: 

      \begin{equation}
      \label{eq:Type-III}
      \begin{split}
      a_3(q)= a_3(0)&-
		      \qdiffc{black}{w}{2}\cdot\twocyclec{red}{uv}                          
      \end{split}
      \end{equation}

      Thus, the topology guarantees the fulfillment of both the necessary \textit{and} the 
      sufficient conditions for the existence of Turing patterns, independently
      of the diffusion rates. 
      Understanding the topological mechanism that underlies the relaxation of diffusion constraints facilitates the design of Turing networks.
      Relaxation occurs if at least one stabilizing $\ell_R$-subgraph of the same 
      size than the destabilizing module has at least one complementary node that is immobile. 
      The immobile node necessarily belongs to the destabilizing module, since its 
      complementary nodes must all be diffusible. It follows that making non-diffusible a node that is complementary to several stabilizing cycles is 
      an efficient way to relax the diffusion constraints.
      Likewise, as more nodes of the destabilizing module are assumed to be immobile,
      it is more likely that subgraphs of the network loose their stabilizing influence and   
      that the diffusion constrains are weakened. 
      This is the reason why in larger networks, which can have
      larger destabilizing modules that accommodate more immobile nodes, the fraction of Type-II and Type-III networks increases. 
      Thus, the graph-based analysis makes the explanation of this observation straightforward.
      \subsection*{Turing filters and Oscillatory Turing networks}
      The extreme case of Turing networks in which all nodes of the destabilizing module are
      immobile deserves special attention. We previously discovered that these networks are all Type-III and their dynamic
      behavior is qualitatively different from standard Turing networks: the wavelength of 
      the emergent pattern is not determined by the network but by the external perturbation \cite{Marcon2016}.     
      
      \begin{figure} [H]
      \centering
      {\includegraphics[width=0.75\linewidth]{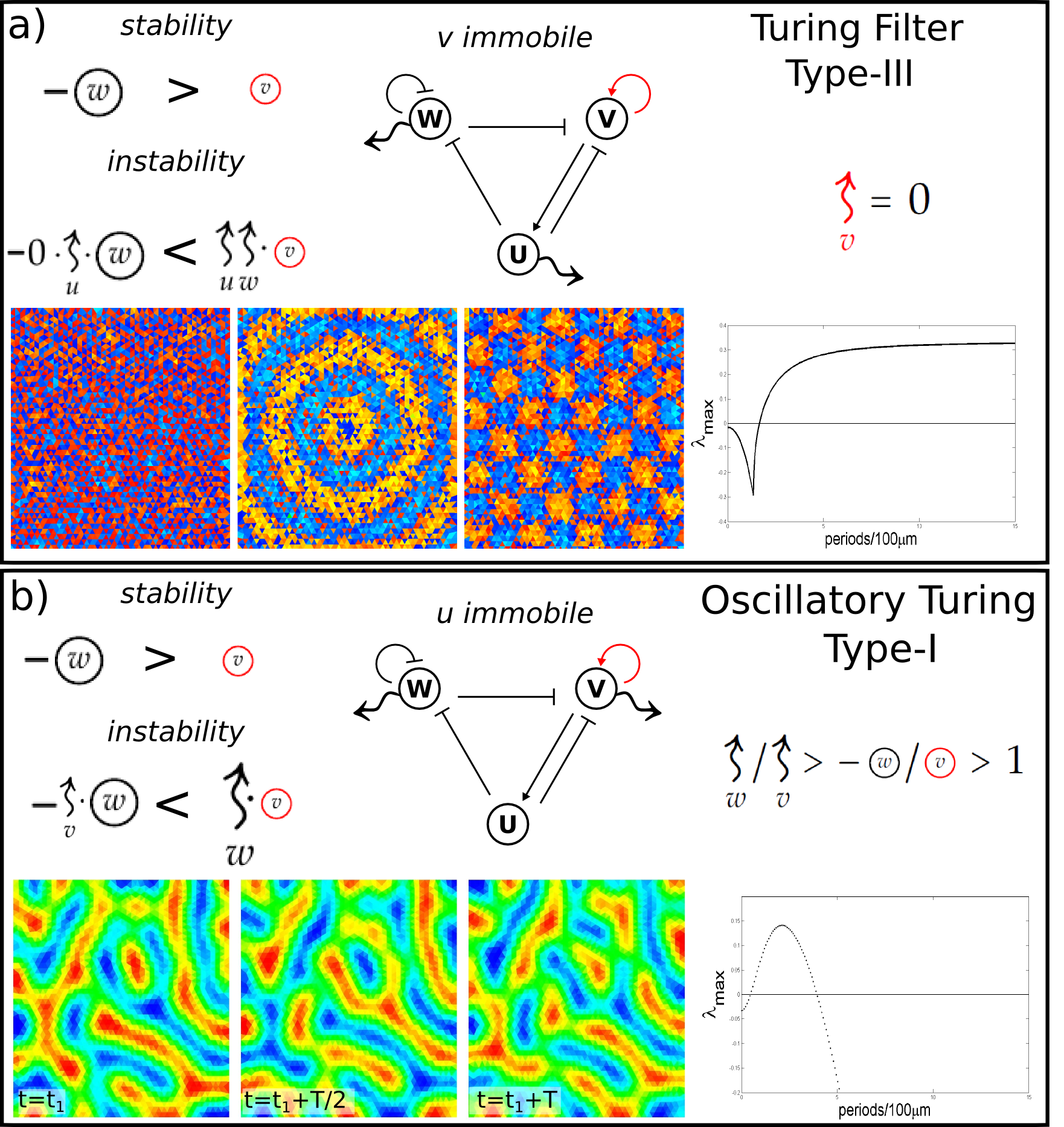}}
      \caption{Minimal topology with 1 immobile node becomes a) Turing filter. The destabilizing module formed by an immobile node.
      The dispersion relationship saturates to a maximum value for infinitely small wavelengths:
      in the presence of noise it amplifies noise, but it can also amplify pre-patterns of wavelength smaller than critical value 
      b) Oscillatory network of Type-I. The dotted dispersion relationship indicates a complex eigenvalue.} 
      \label{fig:fig6}
      \end{figure}
      \vspace{0.25cm}
      The reason is that the dispersion relationship does not have a peak that determines the pattern wavelength. Instead, 
      the maximum eigenvalue grows monotonically from a negative value at $q=0$ and tends asymptotically to a maximum positive 
      value for large wave-numbers. The formal proof of this result is given in SM6 using Rouche's theorem. 
      Thus, modes with a wavenumber bellow the critical value are not amplified, whereas modes with a larger wavenumber grow with 
      comparable speeds. Therefore, the emergent patterns do not have a characteristic wavelength determined by the network. 
      Instead, the initial perturbation that kicks the system out of the homogeneous 
      equilibrium is what determines the pattern that emerges. If the initial 
      perturbation has a spatial structure with a wavelength smaller than the critical value, the system amplifies it to form a 
      stationary pattern with the same spatial structure.  Conversely, an initial pre-pattern with wavelength above the critical value is not amplified.      
      If the homogeneous state is driven out of equilibrium by a small amplitude 
      white noise, all the modes present in the perturbation grow. In this
      scenario, the modes that grow faster are those with infinitely small
      wavelength, and for this reason the system evolves to form a stationary
      salt-and-pepper pattern. In this sense, this subset of Type-III networks are not
      genuine spontaneous pattern forming systems and they could rather be called Turing filters.\\      
      The results obtained so far have focused on stationary Turing patterns.
      However, the analysis can be extended to oscillatory Turing patterns with only minor modifications.          
      Indeed, the classification according to diffusion constraints and the topological arrangements that 
      distinguish the different Types carries over for most networks generating oscillatory Turing patterns.       
      These are the networks in which the instability occurs when a coefficient $a_{k<N}(q)$
      turns negative, while $a_{N}(q)$ remains positive. Hence, the subgraph that causes the 
      instability spans $k<N$ nodes. An important difference is that the conditions for the
      existence of oscillatory Turing patterns are sufficient but not necessary.      
      This means that not all networks capable of generating oscillatory Turing 
      patterns are covered. The networks left out of the analysis are however 
      rare and are subject to severe constraints in their kinetic parameters although, interestingly, they 
      can be built without any positive cycle and do not require differential diffusivity (see SM1 for details).   
      Oscillatory Turing filters also exist and like their stationary counterparts are 
      characterized by having a destabilizing module composed of non-diffusible nodes. 
      They have less patterning power than Stationary Turing filters: noisy perturbation 
      inputs or stochastic dynamics combined with oscillations destroy any pre-pattern and evolve to form an
      an oscillating salt-and-pepper pattern of large amplitude. 
      However, if the system is assumed to follow deterministic dynamics, the amplification of 
      input perturbations with a characteristic wavelength that falls in the flat region of the
      dispersion relationship is comparable to the amplification of salt-and-pepper patterns. 
      In this instance, Oscillatory Turing filters can produce a pattern that results from the 
      oscillatory coupling of several modes and rich dynamics ensue.           
           
           \subsection*{Topology and Robustness}
      A common criticism about Turing systems is that they are not robust, because small parameter  variations impair their patterning potential. 
This feature is related to what has been referred as the fine-tuning problem, noting that Turing systems require either unrealistic separation of
diffusion scales or unphysical fine-tuning of kinetic parameters \cite{Butler2011}. Generally, it is not known what determines the size of the parameter space of
a Turing system. Murray investigated the robustness of several 2-node Turing models and found large variations in the size of their Turing space\cite{Murray1982}.
Several biologically motivated models have shown that the size of the parameter space of receptor-ligand based Turing systems massively increases when the diffusion of the receptor 
is restricted to single cells \cite{Rauch2004,Kurics2014} or is assumed to be immobile \cite{Levine2005,Klika2012}.
Previously, we made a computational screen to find all minimal Turing networks of 3 and 4 nodes with two diffusible species \cite{Marcon2016}. The calculation 
of the size of the parameter space of these networks revealed a trade-off between stability and instability conditions. 
These observations can be partially understood in the framework of our theory. 
All minimal Turing networks of a given number of nodes can be grouped into a limited number of topological families. 
A topological family has a unique and minimal distribution of cycles that allows to build networks that can be stable
without diffusion and that can undergo diffusion-driven instabilities.

        \begin{figure} [h]
      %\captionsetup[subfigure]{labelformat=simple, labelsep=parens}%{labelformat=empty}
      \centering
      {\includegraphics[width=0.75\linewidth]{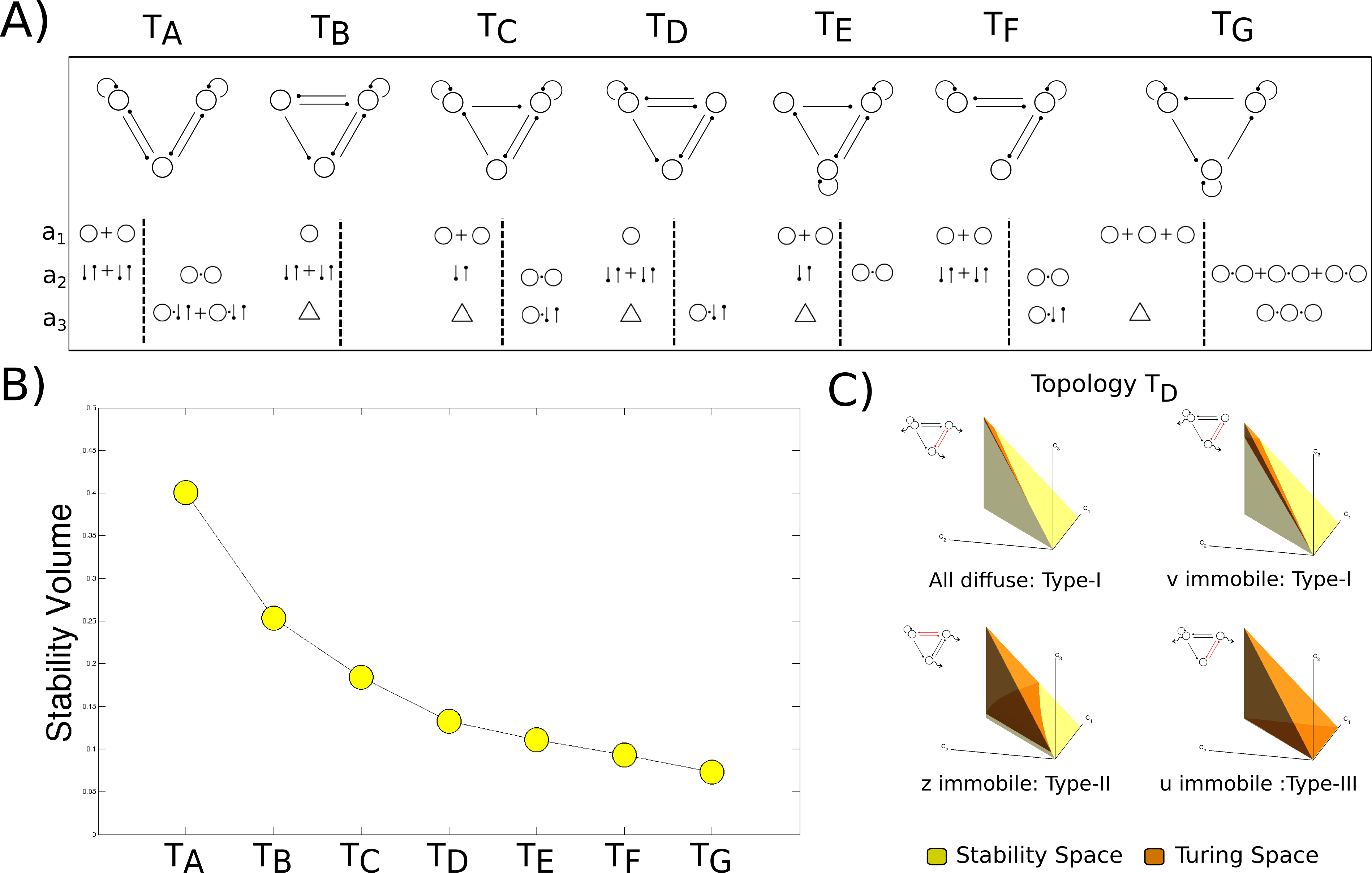}}
      \caption{a) Seven topological families contain all 3-node Turing networks. Unique distribution of cycles and $\ell_R$-subgraphs is shown bellow. 
               b) Stability space of 3-node topological families. The Stability space for all Turing networks within a family is invariant  
               c) Comparison of the Turing space for $T_D$ networks when all nodes diffuse (top-left, fine-tuning necessary) or one of the nodes is immobile (rest, fine-tuning not required) with $d=1.5$ in all cases.}%
      \label{fig:fig4}
      \end{figure}
      
For example, the 21 non-isomorphic Turing networks of 3 nodes found in our previous computational screen \cite{Marcon2016} can be grouped 
into just 7 topological families shown in \ref{fig:fig4}a. 
Similarly, we found 64 non-isomorphic Turing networks of 4 nodes that can be classified into just 12 topological families, which are shown in SM10. 
A conjecture about the relationship between the number of nodes and edges required to build a minimal Turing topology and other relevant
properties are also discussed in SM10. 
Crucially, all networks that belong to the same topological family have an identical Stability space. 
Furthermore, the size of the Stability space of different topological families varies markedly, as shown in \ref{fig:fig4}b.
The reason is that the Stability space is formed by the intersection of the hypersurfaces defined by the Routh-Hurwitz 
stability conditions. Importantly, these conditions depend only on the network cycles and $\ell_R$-subgraphs that they form. 
Because of this, and restricting the analysis to systems in which 
the interactions between species do not depend on the steady state, the Stability space is determined exclusively by the topological
family of the network. 
In turn, the Turing space is the fraction of the Stability space that is compatible with diffusion-driven instabilities and
 is determined by the diffusion rates. Precisely, the key variable is the ratio between the diffusion rates of the nodes
 that induce the destabilizing module and its complementary nodes. If all nodes diffuse, the Turing space tends to zero as 
this ratio tends to one. This is precisely the source of the fine-tuning problem: if realistic differences in diffusion
 rates are assumed, the size Turing space becomes infinitesimal. The fundamental cause of this behavior is the opposite 
requirements of stability without diffusion and the condition stemming from Decartes' rule necessary for Turing instabilities.
  These two requirements can be combined in a particularly simple form if the destabilizing module $\ell^{ins}$ 
 does not overlap with the other stabilizing subgraphs of the same  size ($\ell_1^{st},...,\ell_m^{st}$):     
\begin{equation}
d^{N-k} > \frac{{\left| {\ell _1^{st} } \right| + ... + \left| {\ell _m^{st} } \right|}}{{\left| {\ell ^{ins} } \right|}} > 1
\end{equation}
where $d$ is the ratio between the diffusion of nodes complementary to the destabilizing module and the nodes 
that induce it, and $k$ is the size of the destabilizing module. Note that as $d$ tends to 1 the space of parameters that can 
fulfill both inequalities vanishes.
However, as demonstrated before, this behavior depends on the assumption that all nodes diffuse. 
If there are nodes in the destabilizing module that are immobile, the second inequality does not apply and networks of
different Type according to the diffusion constraints can be obtained.
Even if all the nodes that diffuse do so at the same rate, the Turing space does not vanish, as shown in \ref{fig:fig4}c for a particular topological family.
Thus, the robustness of a Turing network results from a combination of two factors: i) the topological family,
which determines the volume of the Stability space and ii) the Type given by which nodes are immobile, which 
determines the Turing space. 
This illustrates the power of analyzing Turing systems through a topological lens, since it shows that the
fine-tuning problem is not intrinsic to Turing systems, it reveals its source, and how to bypass it.   

\subsection{Topology and Pattern Phases}
The original 2-node network postulated by Turing can be implemented in
      two different forms, typically referred to as activator-inhibitor and 
      substrate-depleted models \cite{Gierer1972}. The activator-inhibitor network forms a periodic 
      pattern in which the concentrations of the two species are in-phase, 
      whereas in the substrate-depleted they are out-of-phase. Both
      networks have the same topology: a node with a positive loop and
      a node with a negative loop connected by negative cycle of length 2,
      but with the signs of the edges flipped. Thus, the two networks 
      have the same distribution of cycles and cycle signs but differ 
      in the signs of their edges, which leads to the difference in patterns.
      For both networks, the analytic expression of the conditions for 
      Turing instability and the dispersion relationship are identical 
      \cite{Miura2004}, so that the wavelength and speed of growth of
      the patterns generated are also identical, provided that the kinetic
      parameters have the same absolute values. The analysis of Turing networks 
      through the graph-theoretical lens shows how these properties carry over 
      for networks of any number of nodes. 
      
          \begin{figure} [h]
      %\captionsetup[subfigure]{labelformat=simple, labelsep=parens}%{labelformat=empty}
      \centering
      {\includegraphics[width=0.75\linewidth]{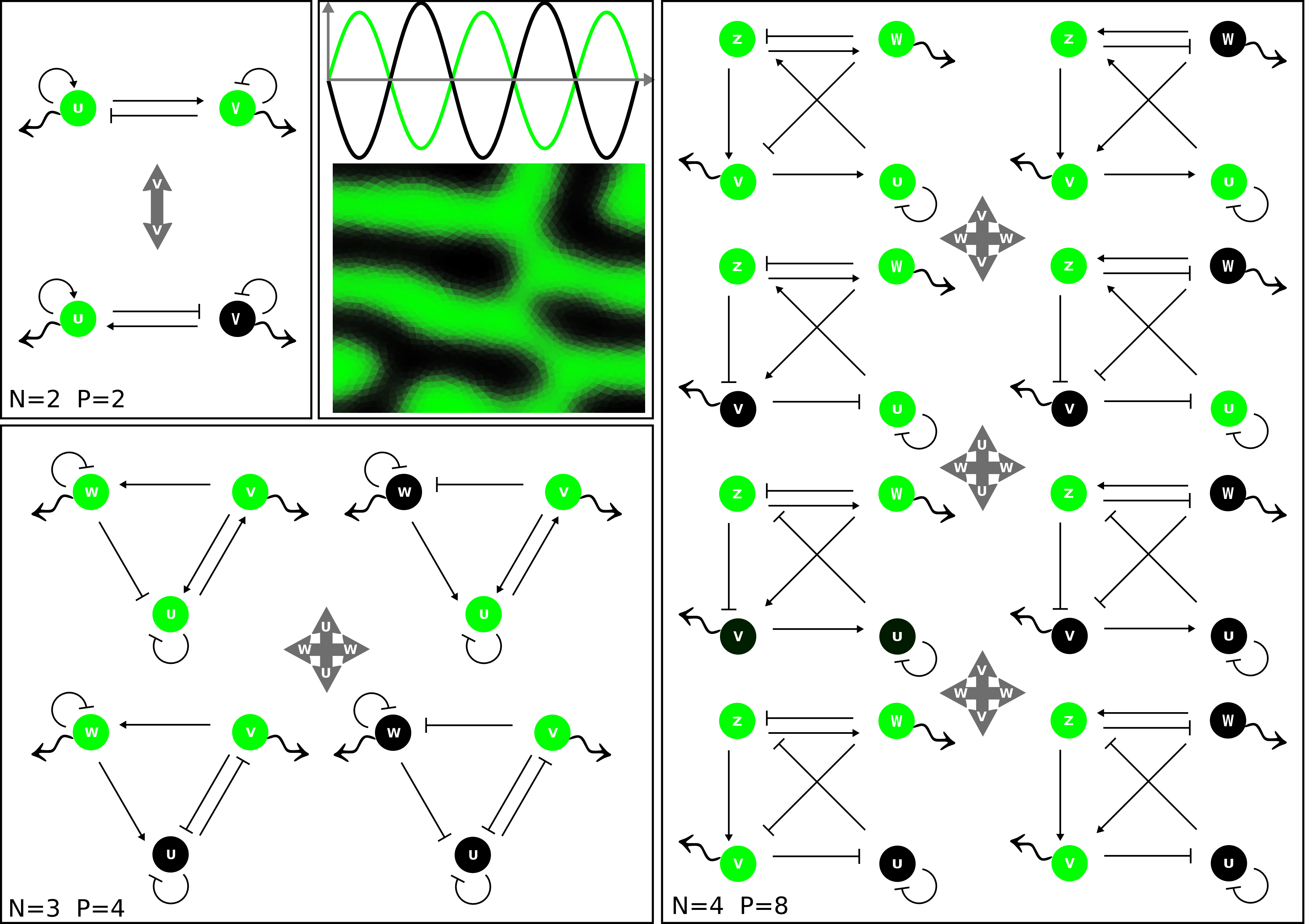}}
      \caption{There are $2^{N-1}$ ways to group $N$ species in a \textit{green} and a \textit{black} phase of Turing pattern. 
      A topology allows to construct $2^{N-1}$ different Turing networks with the sign of the cycles invariant.
      Each network makes one of the patterns. Grey arrows indicate the node to which the sign of outgoing 
      and ingoing edges has been switched, producing a change in phase.} 
      \label{fig:fig8}
      \end{figure}
      
      First, networks that have the same topology and the same
      distribution of cycle signs are restricted to identical requirements 
      for the existence of  diffusion-driven instabilities and generate
      patterns with the same wave-length and growing speed. The reason 
      for this is that the conditions for stability without diffusion 
      and for Turing instability depend exclusively on $\ell$-subgraphs 
      and cycles, rather than individual kinetic parameters. Thus, kinetic parameters 
      and diffusion rates are subjected to the same restrictions for the existence of Turing patterns. 
      For the same reason, the dispersion relationship emerging from the solution of the linearization
      problem \ref{eq:Pq} is identical for networks with the same topology and cycle sign distribution, 
      so that the dynamics and wavelength of the pattern that they generate are the same. 
      
      Second, $N$ species can be grouped in two phases in exactly  $2^{N-1}$ different ways. 
      This is therefore the number of different Turing patterns that $N$ species could hypothetically form.
      For example, a 3-node network can form $4=2^{3-1}$ patterns: $1$ pattern with all species 
      in phase and $3$ patterns with one of the species being out-of-phase with the rest.
      A $4$-node network can form  $8=2^{4-1}$ patterns: $1$ with all species in phase, $4$ with 
      one specie being out-of-phase with the rest and $3$ with a pair of the species out-phase
      with the other pair. The general combinatorial proof is given in SM7. The central finding
      is that each of these $2^{N-1}$ patterns is produced by one of the $2^{N-1}$ Turing networks
      that share the same topology and cycle signs but differ in the sign of individual edges.
      Given a network that produces a pattern with a certain distribution of species amongst the two phases, 
      it is possible construct the network that produces the same pattern but with a single species 
      switched to the other phase by flipping the signs of all the edges coming in and out of the corresponding node. 
      Note that this transformation leaves invariant the sign of all the cycles passing through the node, including the loops.
      Applying the same transformation to several nodes at a time, the associated species switch to    the opposite phase. 
      There are exactly $2^{N-1}$ different networks that can be constructed in this way, each generating one of the $2^{N-1}$ 
      possible Turing patterns. The formal proof in terms of similarity transformations of the Jacobian of the reaction-diffusion
      equations is given in SM7. Intuitively, it can be understood that the effect of switching the signs of the edges going 
      in and out of a node is equivalent to inverting the concentration of this node.
      
      In addition, we find that examination of the topology of a minimal network also allows to predict the phases of the reactants. 
      Typically, this is done calculating numerically the eigenvectors of the linearized system for particular choice of parameter values.
      Conversely, the method based on examining the topology (detailed in SM8) is independent of parameter values and provides an intuitive understanding of Turing dynamics.

\section{Discussion}
In real world systems it is easier to obtain reliable information about the topology of a reaction network 
 than precise quantitative values of the parameters such as reaction rates or diffusion constants. This is especially
 so in biology, because these parameters are generally estimated from in vitro experiments, and yet the real “effective”
 values in vivo are likely to be quite different. Consequently, a theory which allows us to determine properties
 of a Turing system from its topology ideally complements quantitative
 measurements to obtain novel insights into patterning systems. In this work we show that central properties of a Turing system can indeed be understood purely through
 the analysis of its topology. 
First, we tackle the question of the requirements of differential diffusion, and in which ways 
these requirements can be relaxed. A commonly discussed method of relaxing diffusion constraints
has been the concept of “hindered diffusion” through the introduction of an immobile node \cite{Lengyel1991}. 
This has been well known since the discovery of the CIMA reaction, but corresponds to a very 
specific topological change - the immobile node reversibly binds a self-activating node in 
such a way that it effectively slows down the diffusion and is inert to the rest of the network. 
Beyond the original CIMA reaction, this idea has also been implicated in biological systems 
in which receptors bound to the cell membrane or the extracellular matrix play the role of an immobile node that reduces
the diffusion of a ligand \cite{Mueller2013}. However, we discovered that there exist 
alternative ways to achieve similar and even greater relaxation of constraints \cite{Marcon2016}, which do not
correspond to the notion of hindered diffusion. The present theory provides a complete understanding of the relationship between topology and diffusion
constraints, and the general principle for their relaxation. 
Indeed, this principle is general in the sense that it explains the relaxation of diffusion constraints in networks
of any size or number of diffusible nodes.  %, as it is the case of approaches based on symbolic algebra software.
We show that the CIMA reaction and models with the same architecture are networks of Type-II, 
so the relaxation of the constraints is not maximized. Other designs, such as Type III 
circuits, are not working by hindered diffusion, and yet allow far more robust systems.
Furthermore, understanding the principle of relaxation allows us
to identify designs that could be optimal for experimental implementation of Turing patterns in chemical or synthetic biosystems.
For example, in SM4 we show how to construct a 4-node Type-III network with just one immobile node.
Crucially, these types of networks do not suffer the fine-tuning problem, a common criticism of 
the plausibility of Turing patterns in real biological systems \cite{Baker2008}. 
Finally, we addressed the question of which patterns can be generated by a Turing 
system, and in particular, which phase relationships the molecules will have with 
respect to each other. We showed here that inverting the sign of the
interactions of a node with the rest of the network has the effect of switching 
its phase in the pattern. This operation can be performed sequentially on any 
node of the circuit, in this way generating all possible phase combinations of 
the nodes of the system - irrespective of how many nodes there are.\\
In summary, our theory explains the relationship between topology and three fundamental 
properties of Turing systems. Our findings should help to finally 
dispel important objections that have been made against the 
role of Turing patterns in biological development. 
They will also be very powerful for the inference of 
circuits underlying real biological patterns, 
and will be of practical use in the race for the design of the first synthetic Turing biosystem \cite{Scholes2017}.

\bibliographystyle{unsrt}  % {unsrt}
\bibliography{dFree}
\end{document}

\maketitle
\tableofcontents

%%%%%%%%%%%%%%%%%%%%%%%%%%%%%%%%%%%%%%%%%%%%%%%%%%%%%%%%%%%%%%%%%%%%%%%%%%%%%%%%%%%%%%
\section{Conditions for stationary and oscillatory Turing patterns}
%%%%%%%%%%%%%%%%%%%%%%%%%%%%%%%%%%%%%%%%%%%%%%%%%%%%%%%%%%%%%%%%%%%%%%%%%%%%%%%%%%%%%%
\label{first:app}

The necessary and sufficient conditions for all the roots of a polynomial to lie in the left half of 
the complex plane are given by the Routh-Hurwitz criterion \cite[see Chapter 6]{Gantmacher1959}.
For a polynomial $P(\lambda)=\lambda^N+a_1\lambda^{N-1}+...+a_N$, the   Hurwitz matrix is defined in terms of its coefficients as:

\begin{equation}
\label{eq:Hurwtiz}
H = \left[ {\begin{array}{*{20}c}
   {a_1 } \hfill & {a_0 } \hfill & 0 \hfill & 0 \hfill &  \ldots  \hfill & 0 \hfill  \\
   {a_3 } \hfill & {a_2 } \hfill & {a_1 } \hfill & {a_0 } \hfill &  \ldots  \hfill & 0 \hfill  \\
   {a_5 } \hfill & {a_4 } \hfill & {a_3 } \hfill & {a_2 } \hfill &  \ldots  \hfill & 0 \hfill  \\
    \vdots  \hfill &  \vdots  \hfill &  \vdots  \hfill &  \vdots  \hfill &  \ldots  \hfill &  \vdots  \hfill  \\
    \vdots  \hfill &  \vdots  \hfill &  \vdots  \hfill &  \vdots  \hfill &  \ddots  \hfill &  \vdots  \hfill  \\
   0 \hfill & 0 \hfill & 0 \hfill & 0 \hfill & 0 \hfill & {a_N } \hfill  \\
 \end{array} } \right]
\end{equation}

The Routh-Hurwitz criterion states that all roots of the  polynomial have negative real parts if and only if 

\begin{equation}
\label{eq:Hurwitzdets}
\Delta_j = \det [H(j)] > 0\,\,\,\,\,\,\,\,\,\,\,\,\,\,\,\,\,\,\,j = 1,...,N
\end{equation}

where $H(j)$ is the submatrix obtained by taking the first  $j$ rows and columns of the Hurwitz matrix. A polynomial 
that fulfills these conditions is said to be stable. If the conditions are not fulfilled, the number of roots with 
positive real part is given by the number of sign changes in the Routh array \cite[pag 230]{Gantmacher1959}: 

\begin{equation}
\label{eq:Routh}
R = [1,\,\Delta _1 ,\,\frac{{\Delta _2 }}{{\Delta _1 }},\,\frac{{\Delta _3 }}{{\Delta _2 }},...,\,\frac{{\Delta _N }}{{\Delta _{N - 1} }}]
\end{equation}

where the last entry of the Routh array can be simplified to $\Delta_{N}/\Delta_{N-1}=a_N$.\\
A corollary of Routh-Hurwitz criterion is that the positivity of all the $a_k$ coefficients is a necessary condition for stability.

\begin{equation}
\label{eq:Corolary}
a_k  > 0\,\,\,\,\,\,\,\,\,\,\,\,\,\,\,\,\,\,\,k = 1,...,N 
\end{equation}

Conversely, if any of the coefficients is negative there exists at least one root with positive real part 
and the polynomial is unstable. Hence, $a_k<0$ for some $k$ is a sufficient condition for the instability of a polynomial.\\
This theorem allows to derive the conditions for the existence of diffusion-driven instabilities in a 
reaction-diffusion system because they are determined by the zeroes of the characteristic polynomial defined as:

\begin{equation}
\label{eq:PQ}
P_q (\lambda ) = \det [\lambda {\bf{I}} - {\bf{F}}^{{\bf{RD}}} (q)] = \lambda ^N  + a_1 (q)\lambda ^{N - 1}  + ... + a_N (q)
\end{equation}
where ${\bf{F}}^{{\bf{RD}}} (q) = {\bf{J}}^{\bf{R}} (r_o ) - q^2 {\bf{D}}$ has been defined in the main text.
The system is assumed to be stable in the absence of diffusion. This means that for $q=0$ all the 
eigenvalues of $P_q(\lambda)$ are in the left half of the complex plane. Diffusion-driven instability
occurs because $P_q(\lambda)$ loses stability when at least one eigenvalue crosses to the right half
of the complex plane for $q>0$. Note that $P_q(\lambda)$ is a real polynomial and its complex eigenvalues always occur in conjugate pairs.\\
Stationary Turing patterns occur when a single, real eigenvalue becomes positive positive and
are associated with a saddle-done bifurcation. Oscillatory patterns, in turn, occur when a 
single pair of complex conjugate eigenvalues crosses to the right half of the complex plane 
and are associated with a Hopf bifurcation. Next we will show that examination of the signs 
of $a_k(q)$ can allow to determine the type of bifurcation that has occurred.\\ 
If a single real eigenvalue of $P_q(\lambda)$ crosses to the right half of the complex plane, 
then it is necessary that $a_N(q)<0$ for some $q>0$. This follows from the identity:

\begin{equation}
a_N (q) = \det [ - {\bf{F}}^{{\bf{RD}}} (q)] = ( - 1)^N (\lambda _1  \cdot ... \cdot \lambda _N ) 
\end{equation} 

A sufficient condition would require ruling out that other eigenvalues cross to the right hand of the complex plane simultaneously. 
This can be enforced if the rest of the Hurwitz determinants are positive\footnote{More complex dynamics can occur when real and complex eigenvalues 
with positive real parts coexist. Ruling this out requires analyzing all the Hurwitz determinants, and this becomes rapidly 
intractable even for networks of moderate size. The simplified conditions $a_k(q)>0$ for $k<N$ and $a_N(q)<0$ guarantee that 
a single real eigenvalue turns positive, but they can not exclude that a complex pair does as well. 
For networks of size $N\le 3$, these simplified conditions are sufficient to guarantee that only a real eigenvalue turns positive, 
since in this case $\Delta_1=a_1(q)>0$ and $\Delta_2=a_1(q)a_2(q)-a_3(q)>0$. For minimal networks of size $N\leq5$, 
we observe that in most cases they also result in a simple stationary Turing instability. This is possibly due to the dominance of the diagonal 
terms $\Delta_i=a_1\cdot a_2...\cdot a_i$ over the off-diagonal 
terms in the Hurwitz determinants \cite{Clarke1993}, which is more pronounced in minimal networks. 
This precludes instabilities associated to Hurwitz determinants 
of order smaller than $N$ turning negative while $a_k(q)>0$ for $k<N$. 
For non-minimal networks, however, the existence of pairs of complex eigenvalues has to be ruled out on a case by case basis.}, 
so that the sign pattern of the  Hurwitz array is $R=[+,+,...,+,-]$. 
However, a necessary condition for stationary patterns is enough for the purposes of investigating diffusion constraints in 
Turing systems, which is the focus of this work.\\
A pair of complex eigenvalues crosses to the right half of the complex plane,
without a real eigenvalue turning positive simultaneously, if and and only if $a_N(q)>0$ while $\Delta_{N-1}(q)=0$ for some range $q>0$. 
The first condition guarantees that there are no real eigenvalues crossing to the right half of the complex plane, 
and the second condition stems from Orlando's formula \cite{Gantmacher1959,Clarke1993}:
\begin{equation}
\Delta _{N - 1}  = ( - 1)^{N \cdot (N - 1)/2} \prod\limits_{i < k} {(\lambda _i  + \lambda _k )} 
\end{equation}

Deriving analytical conditions from the Hurwitz determinant is impractical
even for moderately sized networks. The condition  derived in \cite{Mincheva2008}
that there exists $a_{k<N}(q)<0$ for $q>0$ while $a_N(q)>0$, is simpler but it is only sufficient. 
This means that it can not identify all oscillatory Turing networks. Particularly, networks
associated to a Hurwitz determinant turning negative while the coefficients $a_k$ remain 
positive are not detected, as remarked in the main text. This behavior occurs if 
for example $a_1(q)\cdot a_2(q)<a_3(q)$. Thus, the kinetic parameters are typically 
subjected to severe nonlinear constrains, but interestingly, these networks do not
require a destabilizing module or differential diffusivity.

%%%%%%%%%%%%%%%%%%%%%%%%%%%%%%%%%%%%%%%%%%%%%%%%%%%%%%%%%%%%%%%%%%%%%%%%%%%%%%%%%%%%%%
\section{Coates graph of a reaction-diffusion system and $\ell$-subgraphs examples} 
\begin{figure} [!h]
\centering
{\includegraphics[width=.7\textwidth]{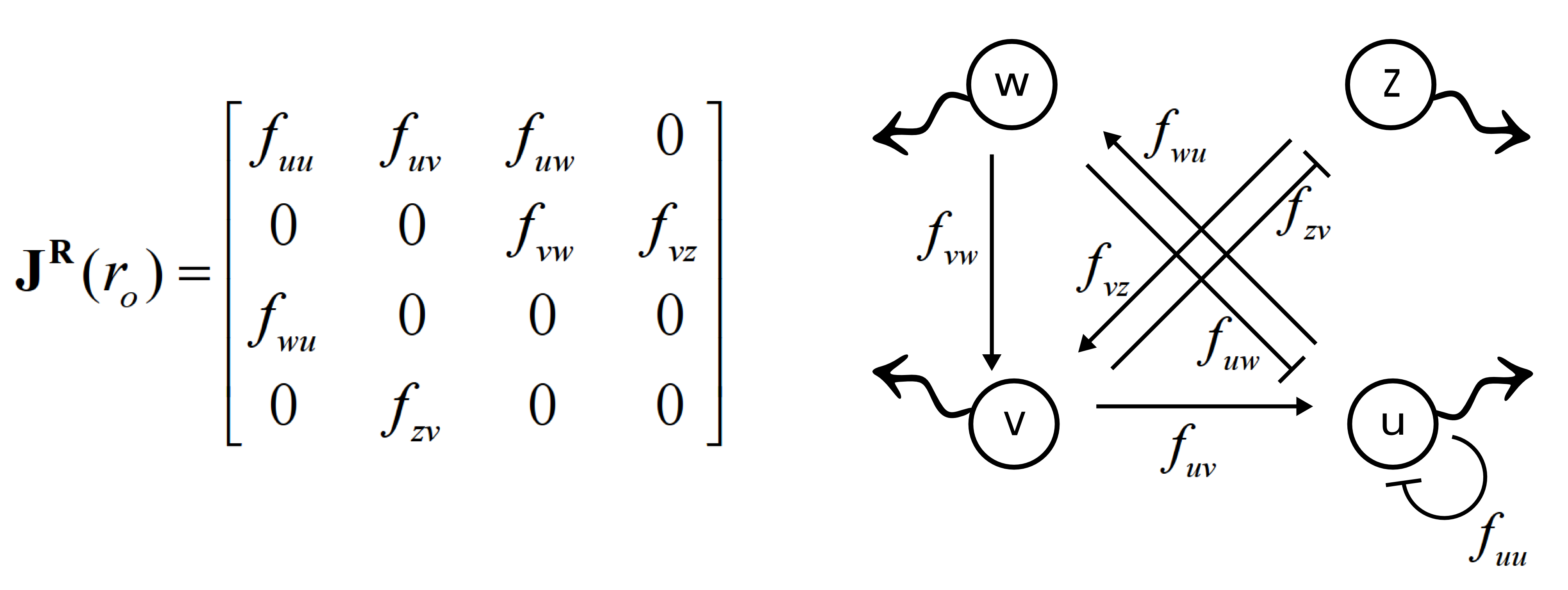}}
\caption{Reaction-diffusion graph of a minimal 4-node network associated to ${\rm{F}}^{{\rm{RD}}} (q) = {\rm{J}}^{\rm{R}} (r_o ) - q^2 {\rm{D}}$. All species are assumed to diffuse.}
\label{fig:Coates}
\end{figure}

%%%%%%%%%%%%%%%%%%%%%%%%%%%%%%%%%%%%%%%%%%%%%%%%%%%%%%%%%%%%%%%%%%%%%%%%%%%%%%%%%%%%%%
\label{second:app}
The reaction-diffusion graph of a reaction-diffusion system has been introduced to illuminate
the relationship between topology and fundamental properties of a Turing systems. The key
element is to associate a weighted digraph to the matrix ${\bf{F}}^{{\bf{RD}}}$ obtained
from the linear approximation of the reaction-diffusion equations. The reaction-diffusion graph will be denoted as $G[{\bf{F}}^{{\bf{RD}}} ]$. Similarly, a graph can be associated only to the jacobian of the reaction term and will denoted as $G[{\bf{J}}^{\bf{R}} ]$. 
The definition of the reaction-diffusion graph is based on the definition of the Coates Graph of a matrix, whose introduction has been attributed to  C.L. Coates \cite{Coates1959}. 
An extensive introduction to this formalism can be found in Brualdi's book \cite{Brualdi08}.
A detailed explanation of the application of this theoretical framework to the analysis of
reaction-diffusion systems was given in \cite{DiegoIniguez2013}. In this section we provide 
a summary of the essential results and examples of $\ell$-subgraphs. The reaction-diffusion
graph of a general $4$ node network is shown in SM fig.\ref{fig:Coates} to illustrate the
definitions given in the main text.

\vspace{0.0cm}

Intuitively, the subgraph induced by a subset of nodes is the subgraph formed by
these nodes and the edges between them. According to this definition, the subgraph 
$I{}_{\gamma _k }$ induced by the nodes $\gamma_k=(i_1,\,i_2,\,...,i_k)$ corresponds
the Coates graph $G[{\rm{F}}_{\gamma _k }^{{\rm{RD}}}]$, where ${\rm{F}}_{\gamma _k }^{{\rm{RD}}}$
is the submatrix formed by row and column indices in $\gamma_k$. In this way we can establish a 
one-to-one correspondence between each principal submatrix   ${\rm{F}}_{\gamma _k }^{{\rm{RD}}}$ and an Induced
subgraph $I{}_{\gamma _k }$.  Examples of this correspondence for two Induced Subgraphs of
the previous network are shown in SM fig.\ref{fig:CoatesII}:

\begin{figure} [!h]
\centering
{\includegraphics[width=.7\textwidth]{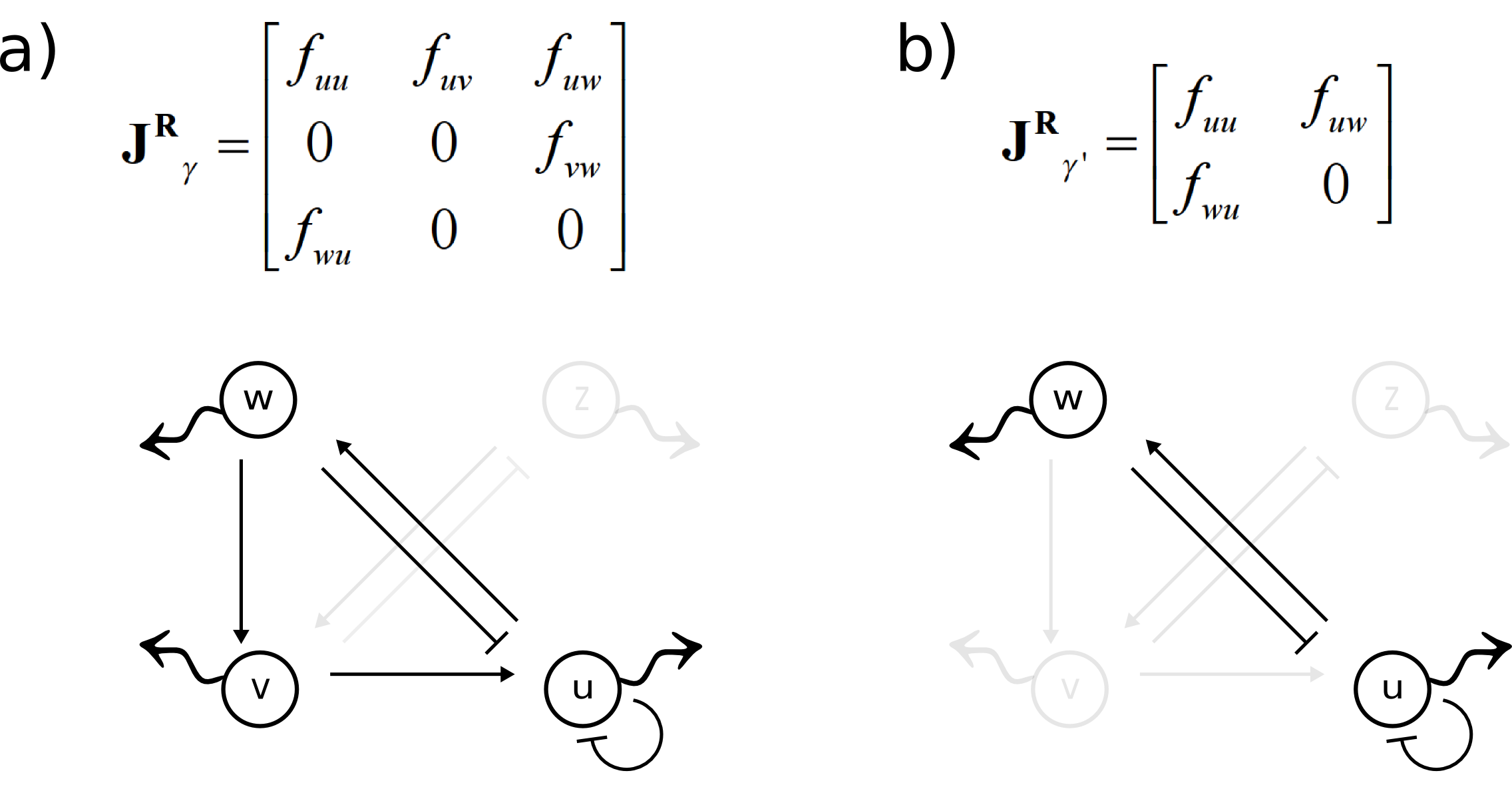}}
\caption{a) Induced Subgraph of size 3 induced by $\gamma=(u,\,v,\,w)$.
b) Induced Subgraph of size 2 induced by $\gamma'=(u,\,w)$. } 
\label{fig:CoatesII}
\end{figure}
\vspace{0.0cm}

There are $\binom{N}{k}$ different $k$-by-$k$ principal submatrices in a $N$-by-$N$ matrix. 
The coefficient $a_k(q)$ of the characteristic polynomial $P_q(\lambda)$ defined in eq.  \ref{eq:PQ} 
can be obtained from the sum of the signed determinants of all the $k$-by-$k$ principal submatrices \cite{Horn1990}:
\begin{equation}
\label{eq:aqII}
{\rm{a}}_k (q) = \sum\limits_{\gamma _k } {( - 1)^k \det [{\rm{F}}_{\gamma _k }^{{\rm{RD}}} (q)]} 
\end{equation}
The Coates formula provides a graphical interpretation of the determinant of a matrix. 
Precisely, the signed determinant of a $k$-by-$k$ matrix ${\rm{F}}_{\gamma _k }^{{\rm{RD}}}$ is 
given by the sum of the weights of all $\ell$-subgraphs contained in the Induced subgraph $I{}_{\gamma _k }$. 
\begin{equation}
\label{eq:Coates}
( - 1)^k \det [{\rm{F}}_{\gamma _k }^{{\rm{RD}}} (q)] = \sum\limits_{\ell  \subseteq I_{\gamma _k } } {w(\ell )} 
\end{equation}
where the weight of an $\ell$-subgraph, as defined in the main text, is given by the product of the cycles 
that it contains and a minus sign for every cycle. Thus, introducing this expression in eq.  \ref{eq:Coates}
we can express $a_k(q)$ as a sum over the $\ell$-subgraphs of size $k$. The example network from SM fig.\ref{fig:Coates} contains
4 cycles that are shown next:

\begin{figure} [!h]
\centering
{\includegraphics[width=.95\textwidth]{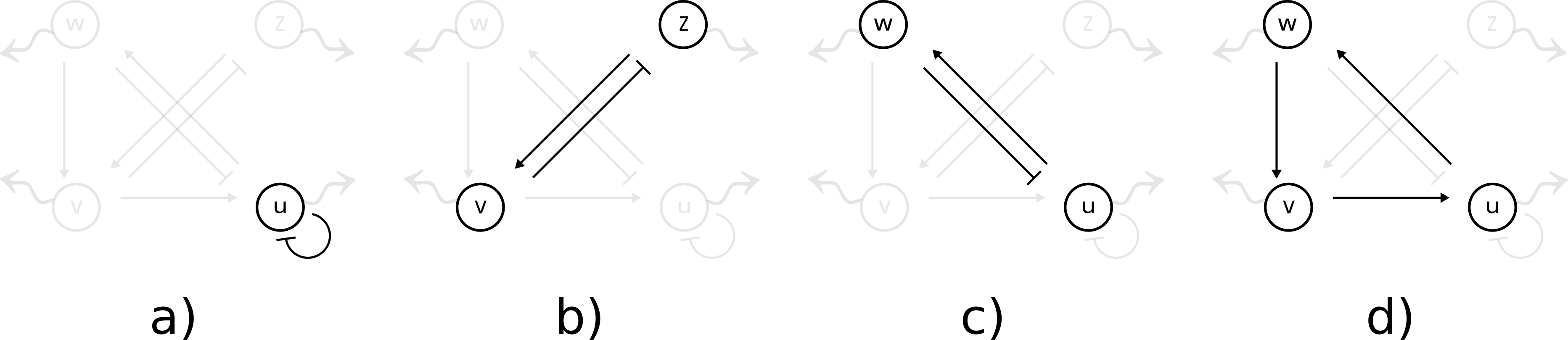}}
\caption{The network from SM fig.\ref{fig:Coates} contains a) one reaction cycle of size 1 b) and 
c) two reaction cycles of size 2 and d) one reaction cycle of size 3.} 
\label{fig:CoatesIII}
\end{figure}
\vspace{0.0cm}

\begin{figure} [!h]
\centering
{\includegraphics[width=.95\textwidth]{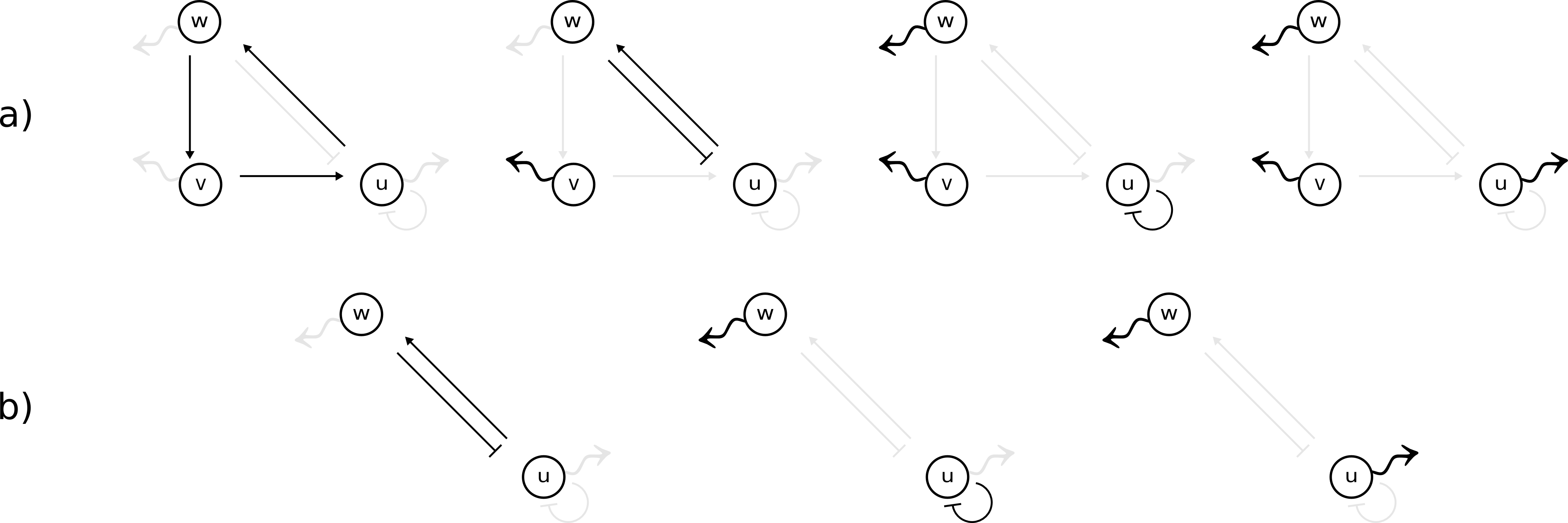}}
\caption{a) $I_{\gamma=(u,\,v,\,w)}$ contains, from left to right: an $\ell_R$-subgraph formed by a cycle of length 3, $\ell_{RD}$-subgraph 
formed by a cycle of length 2 and a diffusive loop, $\ell_{RD}$-subgraph formed by a reaction loop and 
two diffusive loops, an $\ell_D$-subgraph formed by 3 diffusive loops. b) $I_{\gamma'=(u,\,w)}$ contains 
an $\ell_R$-subgraph formed by a cycle of length 2, an $\ell_{RD}$-subgraph formed by a diffusive and a 
reaction loop, an $\ell_{D}$-subgraph formed by two diffusive loops.} 
\label{fig:CoatesIV}
\end{figure}

\begin{figure} [!h]
\centering
{\includegraphics[width=.8\textwidth]{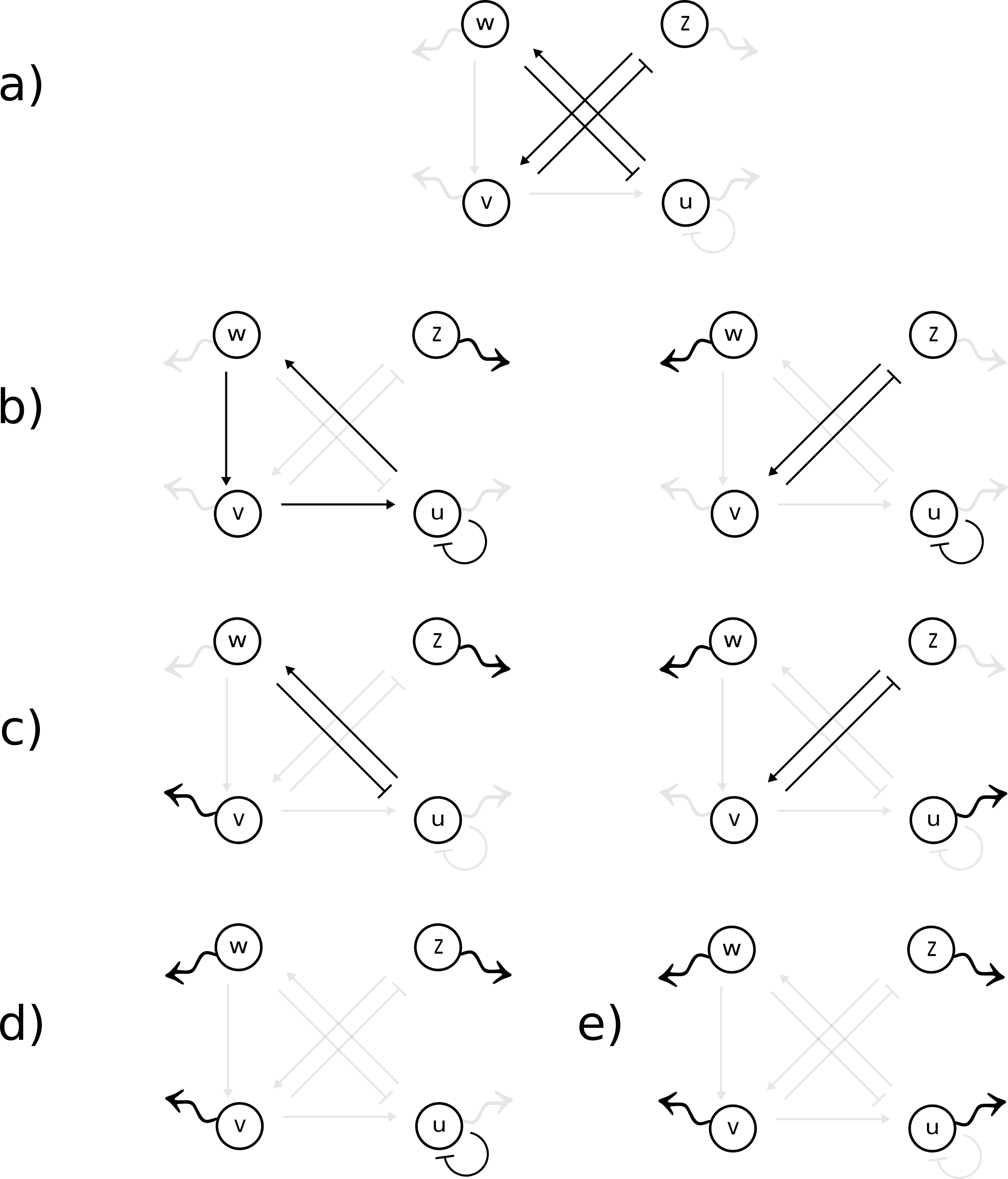}}
\caption{ $\ell$-subgraphs of size 4 contained in the network from SM fig.\ref{fig:Coates} are 
ordered in rows according to the number of diffusive loops contained.} 
\label{fig:CoatesV}
\end{figure}

Non-overlapping combinations of these cycles form the $\ell_{R}$-subgraphs that determine 
the stability of the network without considering diffusion. In turn, combinations of 
these cycles and diffusive loops form the $\ell_{RD}$-subgraphs that determine the patterning dynamics. 
SM fig.\ref{fig:CoatesIV} depicts the $\ell$-subgraphs contained in the examples of
Induced Subgraphs shown previously, and SM fig.\ref{fig:CoatesV} 
shows the $\ell$-subgraphs that span all the nodes in the network.

\vspace{0.0cm}

\newpage

%%%%%%%%%%%%%%%%%%%%%%%%%%%%%%%%%%%%%%%%%%%%%%%%%%%%%%%%%%%%%%%%%%%%%%%%%%%%%%%%%%%%%%
\section{Necessity of a destabilizing module for stationary Turing patterns}
%%%%%%%%%%%%%%%%%%%%%%%%%%%%%%%%%%%%%%%%%%%%%%%%%%%%%%%%%%%%%%%%%%%%%%%%%%%%%%%%%%%%%%
\label{third:app}
A destabilizing module of a reaction-diffusion system has been defined as an Induced
Subgraph of $G[{\bf{J}}^{\bf{R}} (r_o )]$ in which the destabilizing $\ell_R$-subgraphs
outweigh the stabilizing $\ell_R$-subgraphs. In the main text we state that a 
Turing system that generates stationary patterns must have a destabilizing module
of smaller size  than the size of the full system. The necessary condition of 
having a destabilizing module is related to a condition based on the property 
of matrix $s$-stability investigated in \cite{Satnoianu2000} and the minor 
condition investigated in \cite{Wang2001}. These results can both be derived from the 
relationship between $P_0$-matrices and strong stability proved earlier by Cross in \cite[Theorems 1-2]{Cross1978} or 
a similar result by Othmer \cite{Othmer1980}. Here we provide an alternative proof 
that is more direct and clarifies the reason why a destabilizing module (and therefore a positive cycle) 
are necessary for stationary Turing patterns but not for oscillatory Turing patterns. 
We also provide the interpretation of this result in terms of the reaction-diffusion graph.\\ 
 
\textbf{Theorem} \textit{A destabilizing module in a reaction-diffusion system is a necessary condition for the existence of stationary Turing patterns.}\\ 

\textit{Proof:} Let $\mathbf{A}$ be a real $NxN$ matrix and $\mathbf{A}_{\gamma _k }$ denote its $k$-by-$k$ principal submatrix formed by the entries with rows 
and column indices given by $\gamma_k=(i_1,\,...,i_k)$.\\

\textbf{Definition} The matrix $\mathbf{A}$ is said to be a  $P_0$ -matrix if all the signed principal minors are nonnegative:\\
\begin{equation}
\label{eq:Po_mat}
(-1)^k \,\det [\mathbf{A}_{\gamma _k }]\, \geq 0 \qquad   \forall \,\gamma_k \qquad k= 1,...,N 
\end{equation}

The subset $P_0^+$ is formed by the $P_0$ -matrices that have at least one positive signed minor of each order for $k= 1,...,N $. 
The eigenvalues of a $P_0$-matrix are excluded from a wedge around the positive half of the real axis, as proved by Kellog in \cite[theorem 4]{Kellogg1972}.  
Let $\lambda=r\,exp(i\theta)$ be an eigenvalue of a $N x N$ $P_0$-matrix, where $\theta$ is 
the polar angle measured form the positive real axis\footnote{The definitions used in Kellogg's 
paper \cite{Kellogg1972} differ from those adopted here, which follow the definitions given in
Cross paper \cite{Cross1978}. The difference lays in a minus sign in the definition of a $P$, $P_0$ and $P_0^+$-matrices. 
The adaptation of the results to the alternative convention is straightforward, using the following property of the
spectrum of a matrix $\sigma(\mathbf{-A})=-\sigma(\mathbf{A})$}. Then:\\

\textbf{Theorem}
$[$Kellogg$]$ $\lambda=r\,\text{exp}(i\theta)$ is an eigenvalue of a $N x N$ $P_0$-matrix if and only if $\left| \theta  \right| \geqslant \pi /N $\\

Another result, proved by Cross in  \cite[Proposition 1, pag. 257]{Cross1978}, shows that the region of eigenvalue
exclusion of a $P_0$-matrix $\mathbf{A}$ also applies to $\mathbf{A}-\mathbf{D}$, where $\mathbf{D}$ is a nonnegative diagonal matrix:\\

\textbf{Theorem} $[$Cross$]$ If a matrix  $\mathbf{A}\in P_0$ or $\mathbf{A}\in P_0^+$, the same is true of $\mathbf{A}-\mathbf{D}$ for all $\mathbf{D}\geq 0$.\\

 The relevance of this result for the existence of stationary diffusion-driven 
 instabilities is straightforward. Provided that $\mathbf{J}(r_o)$ is a $P_0$-matrix, 
 the eigenvalues of $\mathbf{J}(r_o)-q^2\mathbf{D}$ cannot cross to the positive half
 of the complex plane along the real axis, and stationary Turing pattern can not occur. 
 It follows that is necessary that $\mathbf{J}(r_o)$ is not a $P_o$-matrix. \\
 
\begin{figure} [!h]
\centering
{\includegraphics[width=.8\textwidth]{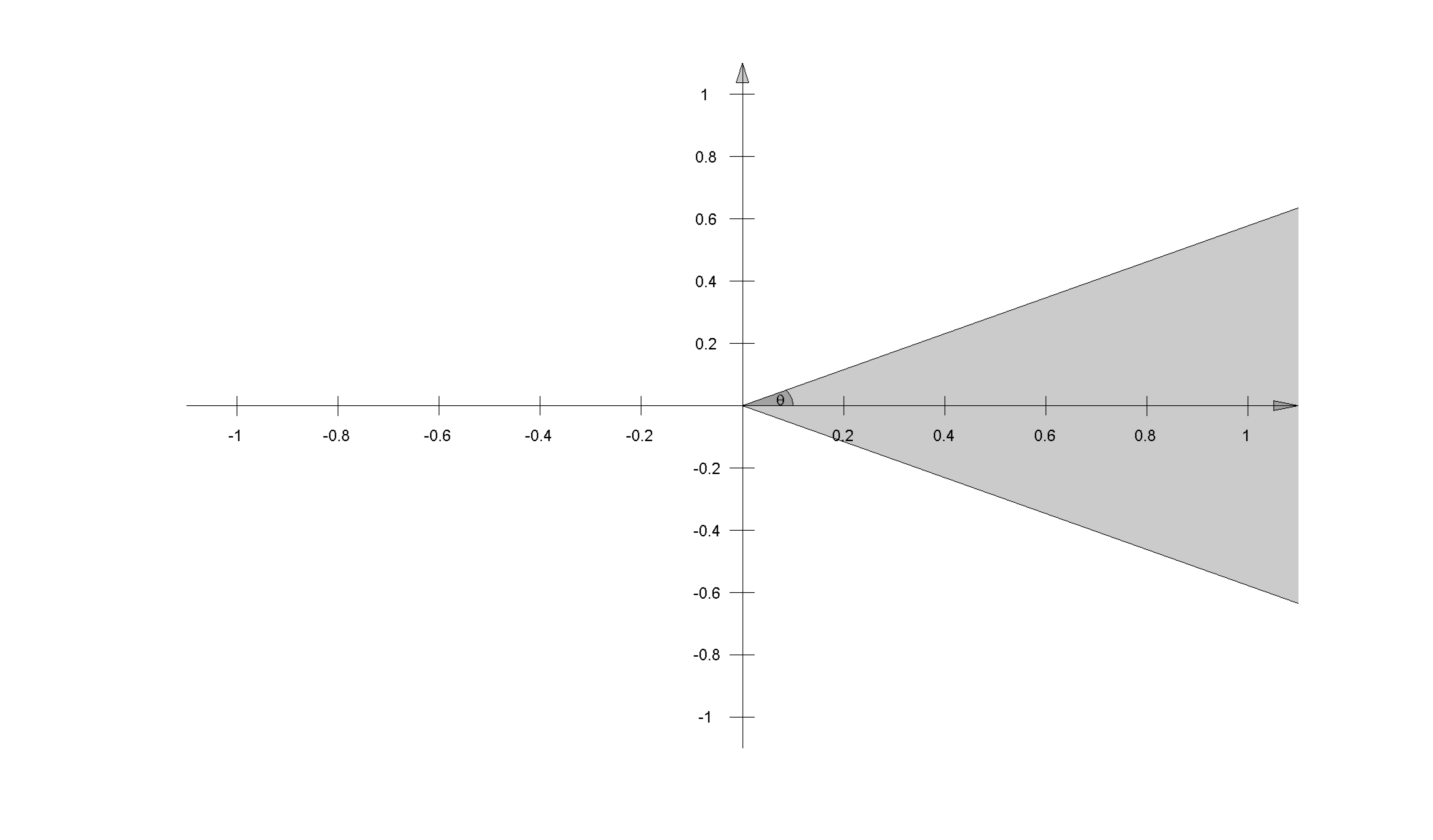}}
\caption{The exclusion region for the eigenvalues of a  $P_0$-matrix $\mathbf{A}$ and $\mathbf{A}-q^2\mathbf{D}$ are the same. $N=6$, $\theta=30^{\circ}$}
\label{fig:Wedge}
\end{figure}
\vspace{0.25cm}

Note that a $P_0$-matrix can have a pair of complex eigenvalues that
cross to the right half of the complex plane with $q>0$. Hence, ${\bf{J}}^{\bf{R}} (r_o ) \notin P_o$ is not necessary 
for oscillatory Turing patterns. This explains the counterexample found in \cite{Satnoianu2005} to a conjecture
given in \cite{Wang2001} that proposes that $P_0$-matrices are strongly stable. This result holds for $N\le 3$
but is not true in general, as was already discussed by Cross in \cite{Cross1978}. These type of 
oscillatory networks are not covered by our theory: the instability is produced when a Hurwitz determinant 
$\Delta_{j<N}$ changes sign while all the coefficients $a_K(q)$ remain positive. Thus, these type of networks can undergo diffusion-driven instabilities 
without having a positive cycle or destabilizing module. However, they seem to be restricted to severe parametric requirements.\\
We now proceed to examine the graphical implications of the necessary condition for stationary Turing patterns.
That $\mathbf{J^R}(r_o)$ is not a $P_o$-matrix implies that there is a principal submatrix of ${\bf{J}}^{\bf{R}} (r_o )$
whose signed minor is negative. Since ${\bf{J}}^{\bf{R}} (r_o )$ is stable, the size of the submatrix is $k<N$. This
submatrix can be associated to the subgraph of $G[\mathbf{J^R}(r_o)]$ induced by a set of nodes $\gamma_k=(i_1,\,...,i_k)$. Further,
the value of the signed principal minor is equal to the weight of the associated induced subgraph, which is in 
turn given by the weights of all the $\ell$-subgraphs contained in it: 

\begin{equation}
( - 1)^k \det [{\bf{J}}^{\bf{R}} (r_o )_{\gamma _k } ] = w(I_{\gamma _k } ) = \sum\limits_{\ell  \subseteq I_{_{\gamma _k } } } {w(\ell )}  < 0
\end{equation}

Therefore, there is an Induced subgraph of size smaller than $N$ in which the destabilizing $\ell$-subgraphs
outweigh the destabilizing $\ell$-subgraph, which completes the proof.

%%%%%%%%%%%%%%%%%%%%%%%%%%%%%%%%%%%%%%%%%%%%%%%%%%%%%%%%%%%%%%%%%%%%%%%%%%%%%%%%%%%%%%
\section{Examples of relaxation of diffusion constraints}
%%%%%%%%%%%%%%%%%%%%%%%%%%%%%%%%%%%%%%%%%%%%%%%%%%%%%%%%%%%%%%%%%%%%%%%%%%%%%%%%%%%%%%
\label{4th:app}
Several networks will be analyzed to illustrate the relaxation of diffusion constraints. 
The first is a 4-node network , this will show how the graph-based analysis is not hampered
by an increase in the number of nodes. The second is a general 3-node Turing network, this
will show that the same is true for non-minimal networks. 
The third is a another 4-node minimal network that, interestingly, becomes a Type-III network
with just one diffusible node. This makes it an ideal design to engineer a system that can generate Turing patterns.
 
\subsection{A 4-node minimal network}
The 4-node network analyzed was shown in SM fig.\ref{fig:Coates}.  The destabilizing module is the
subgraph induced by node $\gamma=(u,\,v,\,w)$ that contains a positive cycle of length 3. First
we will assume that all nodes are diffusible, then we will set one node at a time as non-diffusible,
and finally we will set two nodes as immobile nodes to obtain a Type-III network.
\begin{figure} [!h] 
\centering
{\includegraphics[width=.99\textwidth]{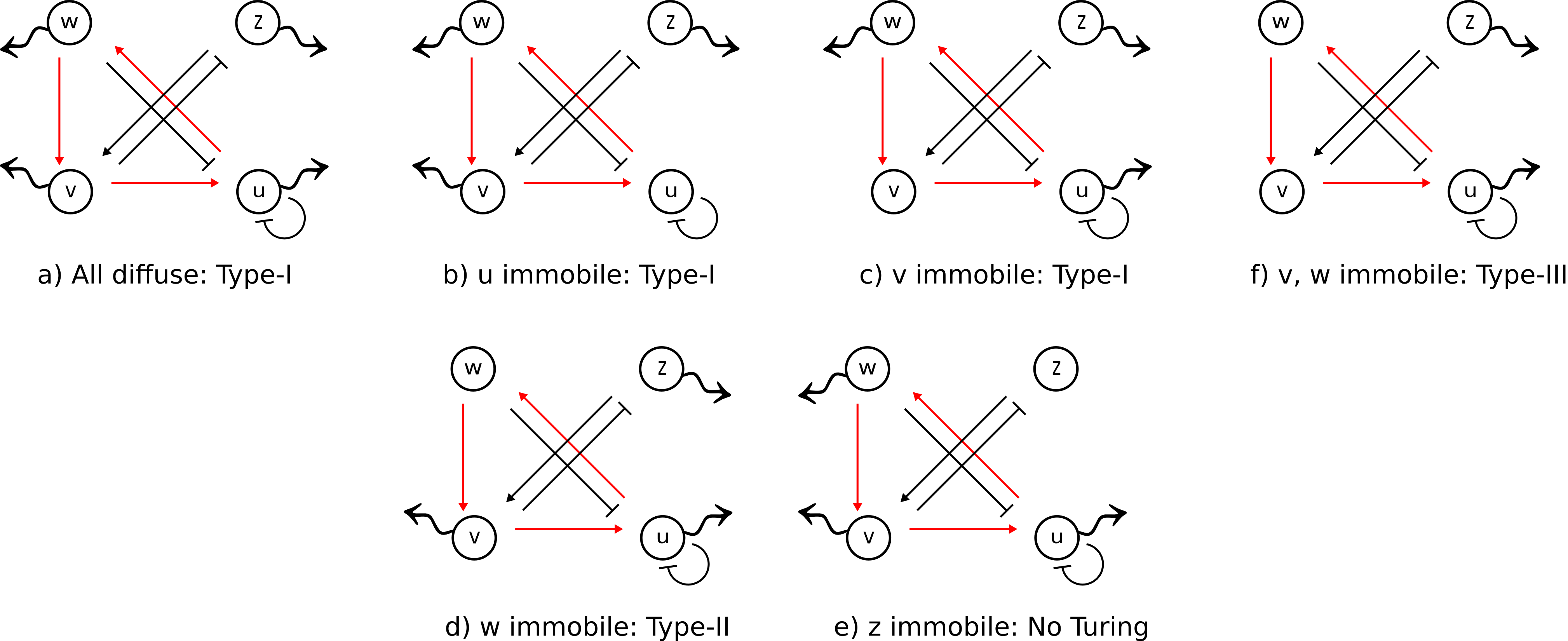}}
\caption{Different degrees of relaxation of the diffusion constrains are achieved by assuming
that different nodes are immobile}
\label{fig:Ralaxation4}
\end{figure}

\subsubsection*{a) All nodes diffuse: Type-I network}
If all nodes are diffusible, the network is necessarily a Type-I Turing network, as demonstrated in the main text. 
The explicit form of the constraints is readily obtain examining $a_3(0)$  and $a_4(q)$. The coefficient $a_3(0)$ 
must be positive for the network to be stable. Conversely, $a_4(q)>0$ for some $q>0$ is required for diffusion-driven instability:
\begin{equation*}
\begin{split}
 a_3(0)&=\qquad\qquad\quad\,\,-\threecyclec{red}{uwv}\pxsp{15}\sloop{u}\cdot\pntwocycle{vz}\xsp{10}>\xsp{10}0\\
 a_4(q)&=+\pntwocycle{vz}\cdot\pntwocycle{uw}
          \pxsp{5}q^2(-\threecyclec{red}{uwv}\cdot\diff{z}\pxsp{5}\sloop{u}\cdot\pntwocycle{vz}\cdot\diff{w}\,) 
          \pxsp{5}q^4(-\pntwocycle{uw}\cdot\diff{v}\diff{z}-\pntwocycle{vz}\cdot\diff{u}\diff{w}\,)
          \mxsp{2}\sloop{u}\cdot q^6\diff{v}\diff{w}\diff{z}
          \pxsp{2} q^8\diff{u}\diff{v}\diff{w}\diff{z}\xsp{1}<\xsp{1}0
\end{split}
\end{equation*} 

The nested structure of the polynomial makes necessary that $\,\difft{z}\,>\,\difft{w}$ so that the 
term of order $q^2$ can be negative and the condition stemming from Decartes rule is fulfilled. 
Hence, the network is of Type-I.

\subsubsection*{b) $u$ immobile: Type-I network}
Setting the diffusion rate of a node to zero does not affect the $\ell_R$-subgraphs formed 
purely by reaction cycles. Hence, $a_3(0)$ is not changed. Conversely, $a_4(q)$ becomes:
\begin{equation*}
 a_4(q)=+\pntwocycle{vz}\cdot\pntwocycle{uw}
          \pxsp{5}q^2(-\threecyclec{red}{uwv}\cdot\diff{z}\pxsp{5}\sloop{u}\cdot\pntwocycle{vz}\cdot\diff{w}\,) 
          \pxsp{5}q^4(-\pntwocycle{uw}\cdot\diff{v}\diff{z})
          \mxsp{5}\sloop{u}\cdot q^6\diff{v}\diff{w}\diff{z}
          \xsp{10}<\xsp{10}0
\end{equation*}

The condition $\,\difft{z}\,>\,\difft{w}$ is still required to make the term of order $q^2$ negative.
The network requires differential diffusivity and is therefore a Type-I network. The topological 
reason that explains why this occurs is that the immobile node $u$ is not complementary to $\sloop{u}\cdot\pntwocycle{vz}$, the stabilizing $\ell$-subgraph of the same size than the destabilizing module.\\
However, setting the node $u$ immobile is not ineffectual: it results in the disappearance of 
stabilizing terms of order $q^4$ and $q^8$. This, in turn, facilitates the fulfillment of the 
condition $a_4(q)<0$ and weakens the requirements on the kinetic parameters. For this reason,
this network has a larger Turing space than the case with all nodes diffusing. 

\subsubsection*{c) $v$ immobile: Type-I network}
In this case, $\,a_4(q)$ and the condition for Turing instability are:
\begin{equation*}
 a_4(q)=+\pntwocycle{vz}\cdot\pntwocycle{uw}
          \pxsp{5}q^2(-\threecyclec{red}{uwv}\cdot\diff{z}\pxsp{5}\sloop{u}\cdot\pntwocycle{vz}\cdot\diff{w}\,) 
          \pxsp{5}q^4(-\pntwocycle{vz}\cdot\diff{u}\diff{w}\,)
          \xsp{10}<\xsp{10}0
\end{equation*}
As in the previous case, the immobile node is not complementary to the stabilizing $\ell$-subgraph 
of the same size than the destabilizing module. Turing instability requires $\,\difft{z}\,>\,\difft{w}$ and
the network is of Type-I. The disappearance of more stabilizing terms of higher order implies that this
network is even more robust than the previous case.

\subsubsection*{d) $w$ immobile: Type-II network}
In this case,$\,a_4(q)$ and the condition for Turing instability are:
\begin{equation*}
 a_4(q)=+\pntwocycle{vz}\cdot\pntwocycle{uw}
          \pxsp{5}q^2(-\threecyclec{red}{uwv}\cdot\diff{z}\,) 
          \pxsp{5}q^4(-\pntwocycle{uw}\cdot\diff{v}\diff{z}\,)
          \xsp{10}<\xsp{10}0
\end{equation*} 
The stabilizing $\ell$-subgraph of the same size than the destabilizing module vanishes. 
Hence, differential diffusivity is no longer required to fulfill the condition stemming from Decartes' rule.
However, the condition $a_4(q)<0$ still involves the diffusion rates, and therefore they are not completely unconstrained. 
The network is then of Type-II. The explicit form of the constraint is: 
\begin{equation}
\frac{\difft{z}}{\difft{v}} \,\,> \,\,-\frac{4\cdot{\pntwocycle{vz} \cdot \pntwocycle{uv}{}^2}}{\threecyclec{red}{uwv}{}^2}
\end{equation}
Note that this inequality can be fulfilled even if $\,\difft{z}\,=\,\difft{w}$ by setting the appropriate values of
the 3 cycles involved in the constraint. An additional degree of freedom is furnished by the cycle that is not 
involved in the inequality. Typically, this results in a larger Turing space than in Type-I networks.

\subsubsection*{e) $z$ immobile: Network cannot produce Turing patterns}
In this case, $\,a_4(q)$ and the condition for Turing instability are:
\begin{equation*}
 a_4(q)=+\pntwocycle{vz}\cdot\pntwocycle{uw}
          \pxsp{5}q^2(+\sloop{u}\cdot\pntwocycle{vz}\cdot\diff{w}\,) 
          \pxsp{5}q^4(-\pntwocycle{vz}\cdot\diff{u}\diff{w}\,)
          \xsp{10}<\xsp{10}0
\end{equation*} 

This inequality cannot be fulfilled, since there is no negative term in $a_4(q)$. The destabilizing
influence of $\threecyclec{red}{uwv}$ is lost because its complementary node is immobile. 
This is true in general: if there are nodes complementary to the destabilizing module that 
are immobile the network cannot undergo stationary Turing instability.

\subsubsection*{f) $w,\,z$ immobile: Type-III network}
If these two nodes are assumed to be immobile, $\,a_4(q)$ and the condition for
Turing instability are:

\begin{equation*}
 a_4(q)=+\pntwocycle{vz}\cdot\pntwocycle{uw}
          \pxsp{5}q^2(-\threecyclec{red}{uwv}\cdot\diff{z}\,) 
          \xsp{10}<\xsp{10}0
\end{equation*} 

The inequality is fulfilled independently of the values of the diffusion rate of $u$ and $z$,
because the destabilizing module is the only contributor to the leading term in $q^2$. Thus, the network is
of Type-III and the Turing volume is equal to the Stability volume.  

\subsection{The complete 3-node network}
\begin{figure} [!h]
\centering
{\includegraphics[width=.99\textwidth]{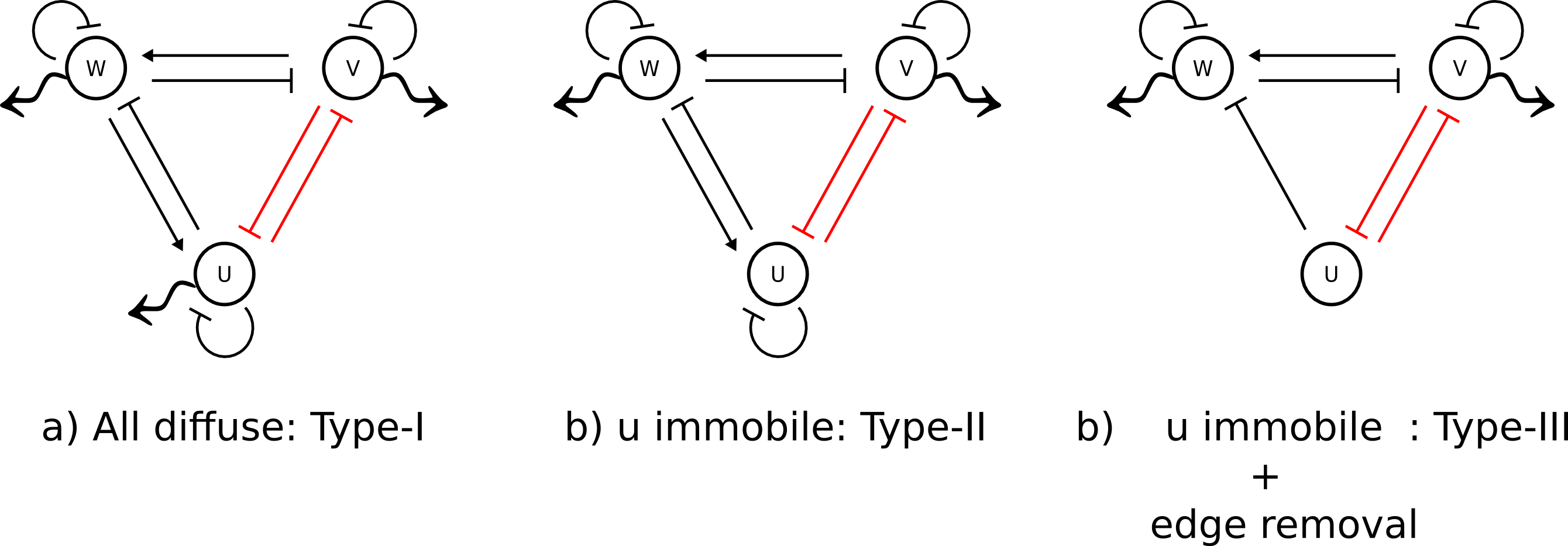}}
\caption{Relaxation of the diffusion constraints in a non-minimal 3N network.}
\label{fig:Relaxation3}
\end{figure}

A 3-node network with all possible interactions between its nodes is shown in SM fig.\ref{fig:Relaxation3}. 
The destabilizing module is assumed to be induced by $u$ and $v$.

\subsubsection*{a) All nodes diffuse: Type-I}
 The coefficient $a_2(0)$ and $a_3(q)$ of the complete $3N$ network are:
\begin{equation*}
\begin{split}
a_2(0)&=\qquad\qquad\,\,\,(\sloop{u}\sloop{v}\mxsp{5}\nntwocyclec{red}{uv})\pxsp{8}(\sloop{u}\sloop{w}\mxsp{8}\pntwocycle{uw})\pxsp{5} (\sloop{v}\sloop{w}\mxsp{5}\pntwocycle{vw})  \\
  a_3(q)&= a_3(0)+\qdiffc{black}{w}{2}(\sloop{u}\sloop{v}\mxsp{5}\nntwocyclec{red}{uv})\pxsp{4}\qdiffc{black}{v}{2}(\sloop{u}\sloop{w}\mxsp{5}\pntwocycle{uw})\pxsp{4}\qdiffc{black}{u}{2}(\sloop{v}\sloop{w}\mxsp{5}\pntwocycle{vw})\\
               &\qquad\qquad-\qdiffc{black}{v}{4}\diffc{black}{w}\cdot\sloop{u}\mxsp{4} 
                 \qdiffc{black}{u}{4}\diffc{black}{w}\cdot\sloop{v}\mxsp{4}
                 \qdiffc{black}{u}{4}\diffc{black}{v}\cdot\sloop{w}\\
           &\qquad\qquad+\qdiffc{black}{u}{6}\diffc{black}{v}\xsp{4}\diffc{black}{w}\\ 
\end{split}
\end{equation*} 
Assuming for simplicity that the diffusion rates of $u$ and $v$ are equal, 
the constraints stemming from stability and Decartes' rule on the
diffusion ratio $d=\difft{w}/\difft{u}=\difft{w}/\difft{v}$ can be combined in the following form:  

\begin{equation}
\label{eq:sType-I}
d\,\, > \,\frac{{(\sloop{u}\sloop{w}\mxsp{5}\pntwocycle{uw})+ (\sloop{v}\sloop{w}\mxsp{5}\pntwocycle{vw})}}{-(\sloop{u}\sloop{v}\mxsp{5}\nntwocyclec{red}{uv})}\,\,>\,\,1
\end{equation}

Thus, the complete 3-node network with all nodes diffusing is a
Type-I network\footnote{Alternatively, if $\sloopc{red}{u}$ were the destabilizing cycle in the network,  
then the coefficient of degree $q^4$ in $a_3(q)$ 
could be negative. Imposing this and that $a_1(0)>0$
and defining the diffusion ratio as before $d$ as before leads to the analogous 
constraint $d\,\, > \,{-(\sloop{v}\pxsp{5}\sloop{w})}/{{\sloopc{red}{u}}} \,\,>\,\,1$. As expected, the constraint of a Type-I network is recovered.}.
\subsubsection*{ b) $u$ or $v$ immobile: Type-II}
The complete 3-node network is transformed into a Type-II network by assuming that either of the two nodes that induce the destabilizing module is immobile. For example, assuming that $u$ is immobile, $a_2(0)$ does not change but $a_3(q)$ becomes:

\begin{equation*}
 \label{eq:sType-II}
\begin{split}
 a_3(q)= a_3(0)&+
                 \qdiffc{black}{w}{2}(\sloop{u}\sloop{v}\mxsp{5}\nntwocyclec{red}{uv})\pxsp{4}
                 \qdiffc{black}{v}{2}(\sloop{u}\sloop{w}\mxsp{5}\pntwocycle{uw})\\
                 &-\qdiffc{black}{v}{4}\diffc{black}{w}\cdot\sloop{u}           
\end{split}
\end{equation*}

In this case,  the conditions stemming form Decartes' rule is fulfilled provided that 
\begin{equation}
\label{eq:constraint1}
d\,\, > \,\frac{{(\sloop{u}\sloop{w}\mxsp{5}\pntwocycle{uw})}}{-(\sloop{u}\sloop{v}\mxsp{5}\nntwocyclec{red}{uv})}
\end{equation}
Note that the diffusion ratio is not unconstrained, but since the expression 
on the right hand side does not need to be bigger than 1, the two diffusible molecules can diffuse at the same rate.  

\subsubsection*{ c) $u$ immobile, removal of 2 edges: Type-III }

To obtain a Type-III network is necessary that the destabilizing module 
is the $\ell_R$-subgraph of smaller order that contributes to $a_N(q)$. 
In this way the destabilizing module forms the leading term in $a_N(q)$. 
The modifications required to achieve this are easily deduced observing 
equation \ref{eq:sType-II}. Thus, removing the loop $\sloop{u}$ and one of the edges from $\pntwocycle{uw}$ the coefficient $a_3(q)$ becomes:

\begin{equation*}
 \label{eq:sType-III}
\begin{split}
 a_3(q)= a_3(0)&-
                 \qdiffc{black}{w}{2}\cdot\nntwocyclec{red}{uv}                          
\end{split}
\end{equation*}

Thus, diffusion-driven instabilities occur independently of the diffusion rates of the diffusible molecules.

\subsection{A 4-node network of Type-III with just one immobile node }

\begin{figure} [!h]
\centering
{\includegraphics[width=.4\textwidth]{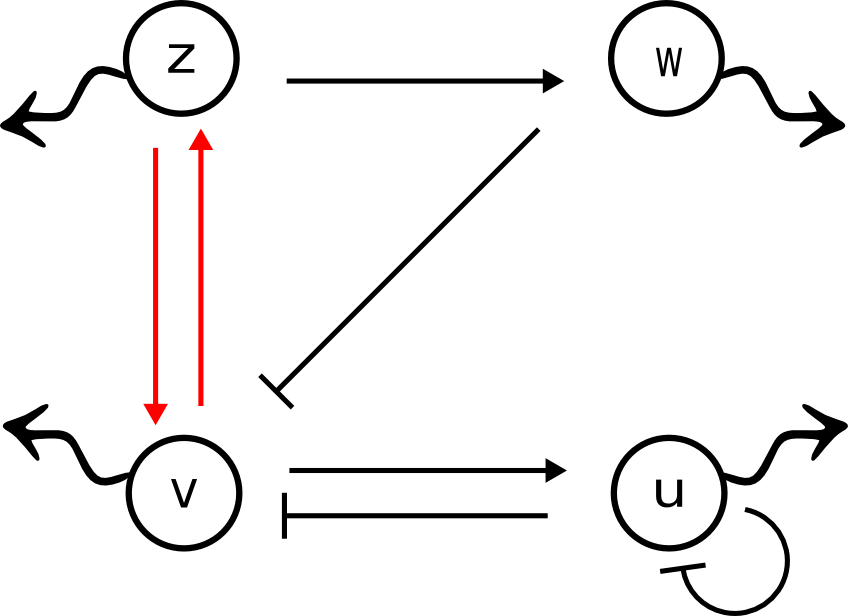}}
\caption{Network that becomes a Type-III assuming $z$ is immobile}
\label{fig:intNet}
\end{figure}

The network shown in SM fig.\ref{fig:intNet} is necessarily a Type-I network
when all nodes are diffusing. The nested structure of the polynomial is the source
of the differential diffusivity requirement , since the $q^2$ and $q^4$ can only
become negative if the diffusion rates of the nodes that form the destabilizing 
module are smaller than the complementary nodes. This can be appreciated in the 
expression of stability conditions $a_2(0)>0$ and $a_3(0)>0$ and the diffusion-driven instability condition $a_4(q)<0$:  

\begin{equation*}
\begin{split}
a_2(0)&=\qquad\qquad\qquad\qquad\qquad\qquad\qquad\quad-\pptwocyclec{red}{vz}\mxsp{25}\pntwocycle{uv} \sxsp{5}{5}{>}0\\
a_3(0)&=\qquad\qquad\quad-\threecycle{vzw}\pxsp{25}\sloop{u}\cdot\pptwocyclec{red}{vz} \sxsp{5}{5}{>}0\\
a_4(q)&=\threecycle{vzw}\cdot\,\sloop{u}
        \mxsp{5}\qdiff{u}{2}\cdot\threecycle{vwz}\pxsp{5}\qdiff{w}{2}\cdot\sloop{u}\cdot\pptwocyclec{red}{vz}
        \mxsp{5}\qdiff{u}{4}\diff{w}\cdot\pptwocyclec{red}{vz}\mxsp{5}\qdiff{w}{4}\diff{z}\cdot\pntwocycle{uv}\\
        &\mxsp{5}\qdiff{v}{6}\xsp{-3}\diff{w}\diff{z}\cdot\sloop{u} 
        \mxsp{5}\qdiff{u}{8}\xsp{-3}\diff{v}\diff{w}\diff{z}\sxsp{5}{5}{<}0
\end{split}
\end{equation*} 

This network is particularly interesting because the destabilizing module is of size two,
which allows to transform it into a Type-III network by assuming that just one nodes, in this case $z$, is immobile.

\vspace{0.25cm}
 The reason is that this change removes the stabilizing effect of all the subgraphs of the same \textit{and} smaller 
 size than the destabilizing module. Hence, the destabilizing module becomes to only contributor to the leading term in $a_4(q)$:

\begin{equation*}
\begin{split}
a_4(q)&=\threecycle{vzw}\cdot\,\sloop{u}
        \mxsp{5}\qdiff{u}{2}\cdot\threecycle{vwz}\pxsp{5}\qdiff{w}{2}\cdot\sloop{u}\cdot\pptwocyclec{red}{vz}
        \mxsp{5}\qdiff{u}{4}\diff{w}\cdot\pptwocyclec{red}{vz}
\end{split}
\end{equation*} 

\begin{figure} [!h]
    \centering
      {\includegraphics[width=0.99\linewidth]{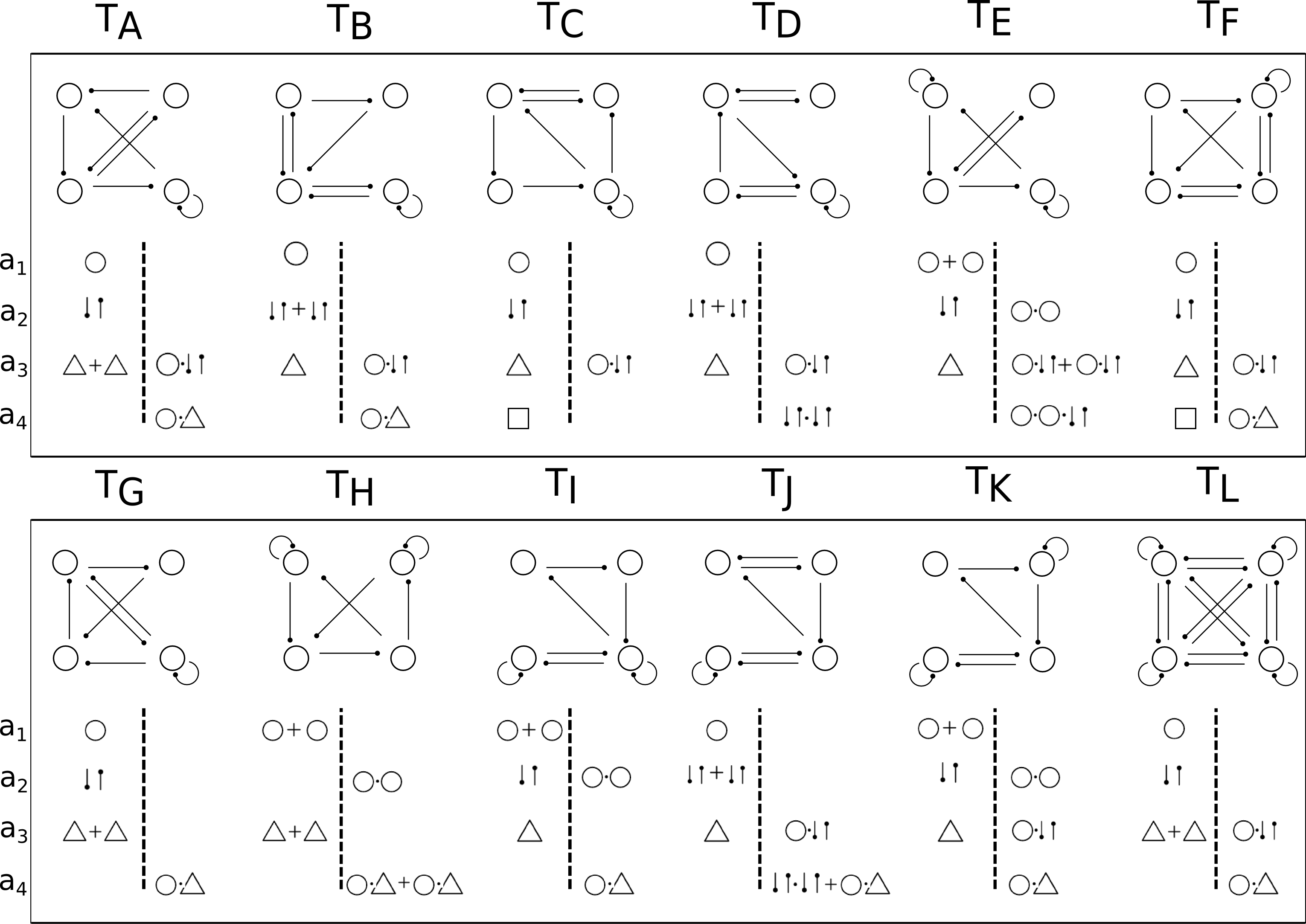}}
      \caption{Turing networks of 4 Nodes can be classified in just 12 topological families.} 
      \label{4nTopologies}
      \end{figure}

Note that this behavior would not be possible if the destabilizing module were of size $3$; in 
this case the cycles of smaller size typically forms a stabilizing term of higher order in $q$
than the destabilizing module. A topology with a positive cycle of size two in turn requires 
that there is another cycle of size two to confer stability. Thus, such an efficient weakening 
of the diffusion constrains with single immobile node is permitted by very specific topological
features. In fact, there are just two $4N$ topological families, listed as $T_B$ and $T_J$ in SM fig.\ref{4nTopologies}, in which 
this behavior is possible. The second topological family, however, generates Oscillatory 
Turing networks for the sets of parameter values, as $a_3(q)$ also turns negative.\\ 
Topology $T_B$ is attractive for the purpose of designing a Turing system experimentally:
it has a small number of edges (minimal in fact), a simple mutual activation cycle (easier to engineer 
than a positive cycle of size 3, be it in a series of chemical reactions or recycling and modifying 
existing signaling pathways in a cell) and just one immobile species (in a chemical system this is
desirable because diffusible reactants should be the norm rather than the exception and in a 
biological system this means that just one transcription factor would be involved). More importantly,
the fact that it is a Type-III network means that the range of values of the kinetic constants 
that fulfill the Turing conditions is very large. Particularly, this design should have a lager 
Turing space than designs that follow the CIMA architecture \cite{Lengyel1992,Horvath2009}.

%%%%%%%%%%%%%%%%%%%%%%%%%%%%%%%%%%%%%%%%%%%%%%%%%%%%%%%%%%%%%%%%%%%%%%%%%%%%%%%%%%%%

\section{Analysis of the relaxation of diffusion constraints in Turing models from the literature}
\label{5th:app}
\subsection{The CIMA reaction }
\begin{figure} [!h] 
\centering
{\includegraphics[width=.8\textwidth]{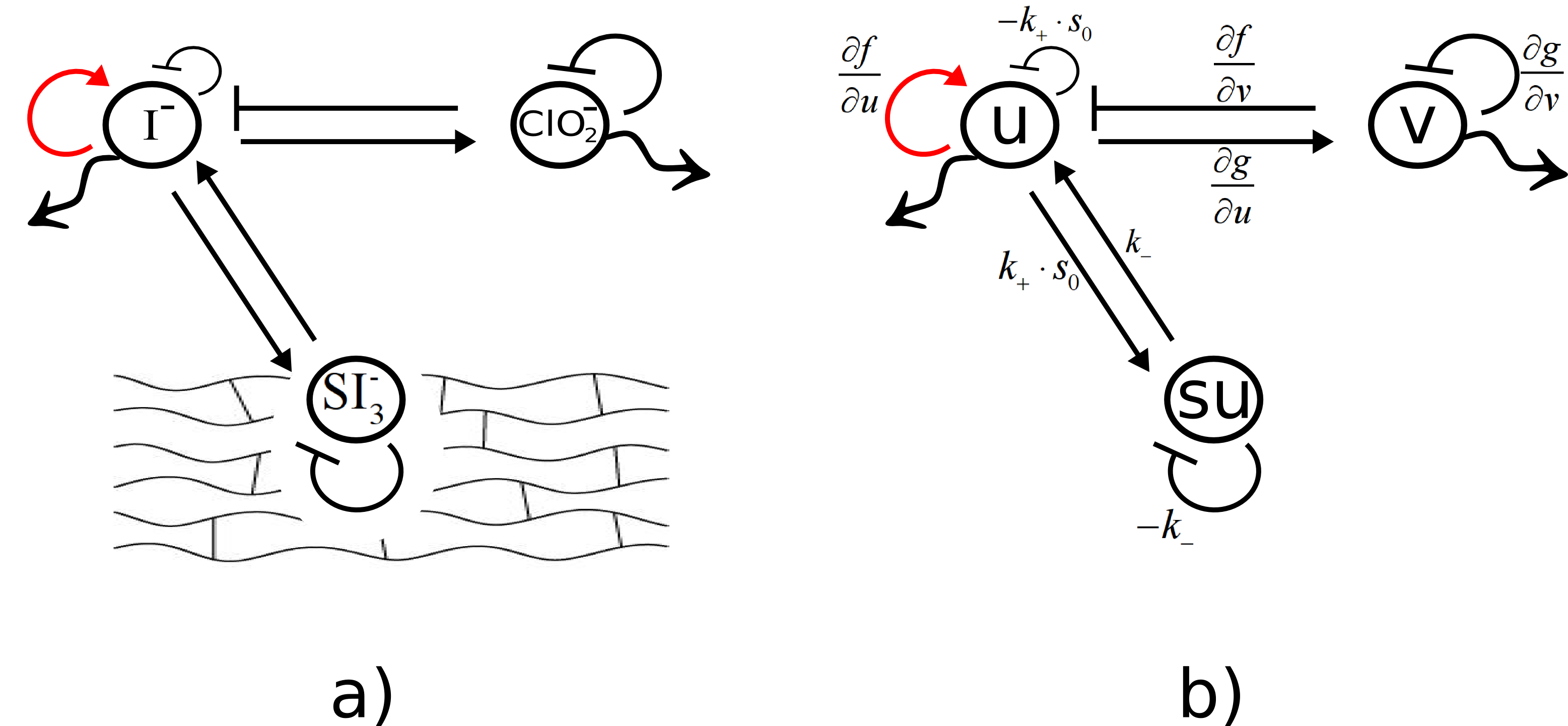}}
\caption{The CIMA model (a) by Lengyel and Epstein\cite{Lengyel1992} and associated reaction-diffusion graph (b). 
Iodine $I^-$ acts as the activator, chlorite $ClO_2^{-}$ as the inhibitor and the starch-iodine complex $SI_3^-$ cannot diffuse. }
\label{fig:CIMA}
\end{figure}

The Chlorite-Iodide-Malonic Acid-starch reaction (CIMA) reaction was the system in which stationary Turing patterns were first observed \cite{Castets1990}. 
Lengyel and Epstein analyzed a model that correctly describes the temporal behavior of the reaction to 
investigate the patterning mechanism underlying the reaction \cite{Lengyel1991}. This analysis suggested 
that starch, introduced as an indicator to visualize the formation of spatial structures, forms a complex
with iodine that cannot diffuse in the gel where the reaction occurred. In this way, the effective
diffusion of the activator (iodine, $I^{-}$) is reduced, producing the difference in diffusion with the 
inhibitor (chlorite, $ClO_2^{-}$) that is required for Turing instabilities. By making a series of reasonable
approximations about the underlying chemical processes, the description of the CIMA reaction can be reduced to 
a 3-species reaction-diffusion model \cite{Lengyel1992}. Writing the concentration of the activator as  $[I^-]=u$, 
the inhibitor $[ClO_2^{-}]=v$, starch  as $s_0$ and the starch-iodine complex $[SI_3^-]=su$, the Lengyel-Epstein model is:
\begin{equation}
\begin{split}
\frac{{\partial u}}{{\partial t}} &= f(u,v) - k_ + s_o  \cdot u + k_ -   \cdot su + D_u \frac{{\partial ^2 u}}{{\partial x^2 }}\\
\frac{{\partial v}}{{\partial t}} &= g(u,v) + D_v \frac{{\partial ^2 v}}{{\partial x^2 }}\\
\frac{{\partial su}}{{\partial t}} &= k_ +  s_o  \cdot u - k_ -   \cdot su
\label{eq:CIMA}
\end{split}
\end{equation} 
The kinetic constants $k_+$ and $k_-$ give the rates of formation and dissociation of the starch-iodine complex.
Lengyel and Epstein further simplified the analysis by assuming that the 
formation and dissociation of the complex is very fast. With this approximation, the concentration of the 
complex is given by $sx=(k_+\cdot s_0/k_-)u$ because it is in instantaneous equilibrium with the activator $u$. The
CIMA model is then simplified to a set of two reaction-diffusion equations for $u$ and $v$ and, importantly, this 
transformation introduces a timescale separation between the two species that reflects that the activator is being
trapped and released by the starch in the medium. Using realistic values for the parameters in the model, Lengyel and
Epstein estimated that without starch the diffusion rate of $v$ ought to be 10 times faster than that of $u$ to generate
Turing instabilities. According to the Stokes-Einstein law, the diffusion of two ions of similar sizes in aqueous solution
cannot possibly be that different. However, the introduction of starch reduces the necessary ratio of diffusion rates to a more plausible 
value of 1.5.\\
Next we use our graph-based formalism to analyze the 
mechanism of relation of diffusion constraints in the CIMA reaction. 
The analysis does not require the assumption of fast complex formation 
and in this respect is more general.
The reaction-diffusion graph associated to the Lengyel-Epstein model of the
CIMA reaction given by eq.\ref{eq:CIMA} is obtained following the procedure 
detailed in the main text and shown in \ref{fig:CIMA}. 

\vspace{0.25cm}

The activator has two loops that correspond to the self-activation
term $\partial f/\partial u$ and the term $k_+\cdot s_0$ that gives the rate of decrease in concentration of $u$ through 
complex formation. In graph notation this is represented by $\sloop{u}=\sloopc{red}{u_a}+\sloop{u_c}$, with the first 
loop accounting for self-activation and the second for complex formation. In addition, the edges that form the cycle 
between $u$ and $su$ and their loops are not independent because the number of iodide molecules is 
conserved and therefore $\pptwocycle{u \cdot su}=-\sloop{u_c}\cdot-\sloop{su}$. These identities lead 
to following simplification of the subgraph induced by $u$ and $su$:  

\begin{equation*}
\begin{split}
(\sloop{u}\sloop{su}-\pptwocycle{u\cdot su})&=[(\sloopc{red}{u_a}+\sloop{u_c})\sloop{su}-(-\sloop{u_c}\cdot -\sloop{su})]\\
&=\sloopc{red}{u_a}\sloop{su}
\end{split}
\end{equation*}

In turn, this simplifies the coefficient $a_3$ of the characteristic polynomial to:
\begin{equation*}
\begin{split}
a_3(q)=& a_3(0)+q^2[\sloop{su}\sloop{v}\diff{u}+(\sloop{u}\sloop{su}-\pptwocycle{u\cdot su})\diff{v}]-q^ 4\sloop{su}\diff{u}\diff{v}\\
      =& a_3(0)+q^2[\sloopc{black}{v}\diff{u}+\sloopc{red}{u_a}\diff{v}]\sloopc{black}{su}-q^ 4\sloopc{black}{su}\diff{u}\diff{v}
\end{split}
\end{equation*}

Hence, the only term in $a_3(q)$ that can be negative in order to fulfill the necessary 
condition for stationary Turing patterns is the coefficient of $q^2$. The form of $a_3(q)$ is 
characteristic of a Type-II Turing system according to the diffusion constrains. Indeed, defining $d=\difft{v}/\difft{u}$, this condition
can be expressed as:
\begin{equation*}
d\,>\,-\frac{\sloop{v}}{\sloopc{red}{u_a}}
\end{equation*}

Importantly, stability requirements do not impose that $\sloopc{red}{u_a}$ is smaller than $\sloop{v}$, because
the network contains other stabilizing loops: 

\begin{equation*}
a_1(0)=-(\sloop{su}+\sloopc{red}{u_a}+\sloop{u_c}+\sloop{v})\,>\,0 \not\Rightarrow \sloopc{red}{u_a} < -\sloop{v}
\end{equation*}
Hence, the diffusion ratio can be equal to 1 and even smaller, depending on the parameter values of $\sloopc{red}{u_a}$ and $\sloop{v}$. 
The explicit form of the constraints for the kinetic parameters can be obtained examining the rest of the stability conditions $a_i(0)>0$, 
but they do not further limit the diffusion ratio of the activator and the inhibitor.
Thus, this analysis allows us to easily derive the diffusion constraints of the CIMA reaction and does not require the approximation that complex 
formation is very fast. Importantly, all networks designs based on the CIMA architecture  result in a Type-II Turing system, due to the existence 
of stabilizing terms involving more diffusion loops. Examples of such networks can be found in \cite{Rauch2004,Levine2005,Korvasova2015}.

\subsection{Turing models from the literature }
\subsubsection{Models based on the CIMA architecture}
\begin{figure} [!h] 
\centering
{\includegraphics[width=.9\textwidth]{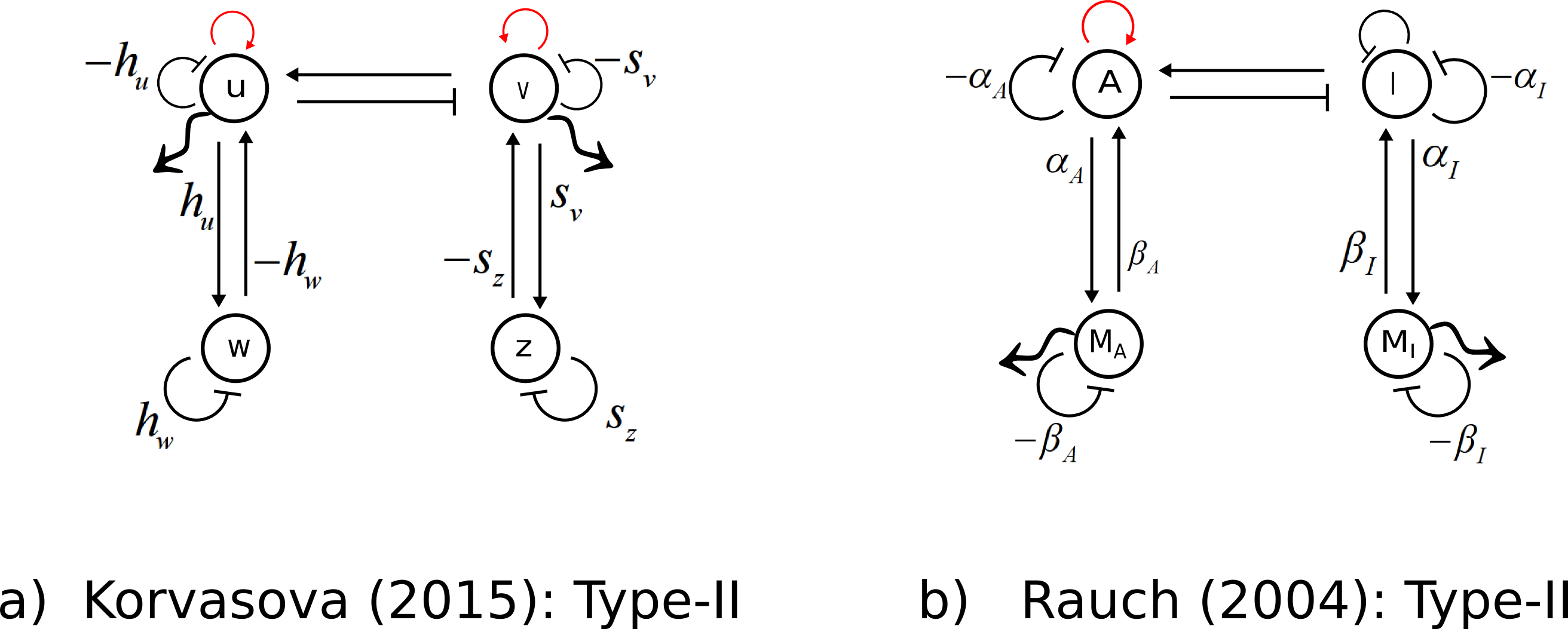}}
\caption{a) Network composed of diffusible activators $u$ and $v$ that bind to immobile substrates \cite{Korvasova2015} b)
A model of biological patterning based on morphogens $M_A$ and $M_I$ that diffuse in the extracellular space and gene 
products that are confined inside cells \cite{Rauch2004}  }
\label{fig:CIMAin}
\end{figure}

In this section we analyze two Turing models inspired by the CIMA reaction. The first is a recent investigation of the conditions
for diffusion-driven instability in the presence of binding immobile substrates by Korvasova and collaborators \cite{Korvasova2015}. 
This is a purely theoretical study that aimed to weaken the restrictive conditions that apply to diffusion rates and kinetic parameters of
2-node Turing networks. To that end, they analyzed the Lengyel-Epstein model \cite{Lengyel1992} of the CIMA reaction using standard linear
stability analysis and algebraic manipulations. They also proposed a 4-node generalization of the CIMA model in which two self-activators bind
two immobile substrates. The corresponding reaction-diffusion graph is shown in SM fig.\ref{fig:CIMAin}a.  
\vspace{0.25cm}
The principle behind the relaxation of diffusion constraints is the same than in the CIMA recation. The transient bonds
between activator and substrates do not change the number of molecules. As in the CIMA model, the subgraphs 
induced by these pairs simplify to $\sloop{u}\sloop{w}-\pptwocycle{uw}=\sloopc{red}{u_a}\sloop{w}$ and $\sloop{v}\sloop{z}-\pptwocycle{vz}=\sloopc{red}{v_a}\sloop{z}$, where the loop notation makes explicit the contribution from self-interaction and complex formation as in the CIMA reaction. The Turing condition $a_4(q)<0$ can then be expressed as:
\begin{equation*}
\begin{split}
a_4(q)=& a_4(0)+q^2 [ -\sloop{z}(\sloop{u}\sloop{w}-\pptwocycle{uw})\diff{v}-\sloop{w}(\sloop{v}\sloop{z}-\pptwocycle{vz})\diff{u}]+ q^4\diff{v}\diff{u}\sloop{w}\sloop{z}\\
=&a_4(0)-q^2 ( \sloopc{red}{u_a}\diff{v}+\sloopc{red}{v_a}\diff{u})\,\sloop{w}\sloop{z}+ q^4\diff{v}\diff{u}\sloop{w}\sloop{z}\\
\end{split}
\end{equation*}
The existence of two activators introduces additional relaxation of constraints  compared to the original CIMA model: 
the necessary condition stemming from Decartes rule of signs is guaranteed because there is a term in $a_4(q)$ formed only by 
destabilizing terms. However, there is a stabilizing subgraph that contains more diffusive loops and because of this the conditions
for diffusion-driven instabilities are not independent of the diffusion rates. Still, as in the CIMA reaction, the diffusion rates can be equal and the network is of Type-II.\\
The second model to be analyzed was proposed by Rauch and collaborators \cite{Rauch2004} as plausible mechanism
of biological pattern formation. The model consists of biochemical reactions between gene products that are 
confined inside cells and therefore can be considered non-diffusible and messenger molecules that are secreted
by cells and can diffuse between them. The associated reaction-graph is shown in SM fig.\ref{fig:CIMAin}b. 
Taking advantage of the simplifications that follow from mass conservation as in the previous model, the $a_4(q)$ coefficient can be expressed as:
\begin{equation*}
a_4(q)= a_4(0)-q^2 [\diff{M_A}\sloop{M_I}(\sloopc{red}{A_a}\sloop{I}-\pntwocycle{AI})+\diff{M_I}\sloop{M_A}(\sloopc{black}{I_a}\sloop{A}-\pntwocycle{AI})]+q^4\diff{M_A}\,\diff{M_I}(\sloop{A}\sloop{I}-\pntwocycle{AI})\\
\end{equation*}

Again, because the nodes $A$ and $I$ are not diffusible there are stabilizing subgraphs that vanish from the 
coefficient of degree $q^2$ in $a_4(q)$. The structure of $a_4(q)$ shows that this network, like CIMA, is also a Type-II.
In this way the model shows how realistic physiological model can result in a Turing system that does not require differential diffusivity.

\subsubsection{Hair follicle formation}
A recent work by Klika et al. \cite{Klika2012} investigates the influence of interactions mediated by 
immobile receptors in a model of hair-follicle patterning in vertebrate skin.  The model was originally
proposed by Mou and collaborators \cite{Mou2006} and is based in their experimental analysis of the interactions 
between three key players in hair follicle patterning: a non-diffusing receptor (Edar), a connective tissue growth 
factor (CTGF) and Bone moprhogenetic factor (BMP). 

\begin{figure} [!h] 
\centering
{\includegraphics[width=.9\textwidth]{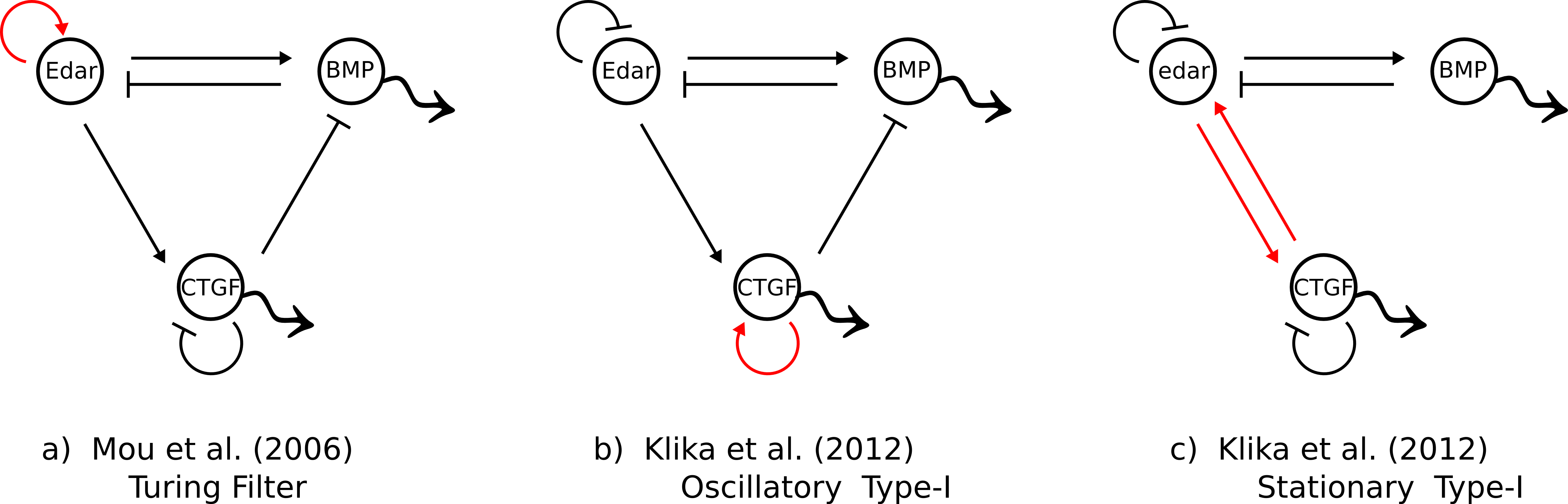}}
\caption{Three alternative models of Hair follicle pattering in proposed in  \cite{Mou2006}(a) and \cite{Klika2012} b) and c). They form a Turing Filter, a Oscillatory Turing networks of Type-I and a Stationary Turing system of Type-I respectively.}
\label{fig:Hair}
\end{figure}
\vspace{0.25cm}

The experimental observations of Mou et al.\cite{Mou2006} led them to propose the network shown in SM fig.\ref{fig:Hair}a. This network 
is a Turing filter. We can reach this conclusion without further analysis because the destabilizing module $\sloopc{red}{Edar}>0$ is non-diffusible. 
The proof of the generality of this result is given in section \ref{5th:app}. Indeed, this can be confirmed examining the form $a_3(q)$:
\begin{equation*}
a_3(q)= a_3(0)+ q^2\diff{c}(\sloopc{red}{e}\sloop{b}-\pntwocycle{eb})+q^2\diff{b}\sloopc{red}{e}\sloop{c}-\sloopc{red}{e}\cdot q^4\diff{b}\diff{c}
\end{equation*}
With the assumed distribution of cycle signs, the condition $a_3(q)<0$ for Turing instability is fulfilled
independently of parameter values for $q$ above a critical value. \\
Klika and collaborators \cite{Klika2012} introduced a series of changes to the model shown in SM fig.\ref{fig:Hair}b due to 
theoretical objections. They postulated that Turing systems must be stable for very large wavenumbers. If this is not
the case, they observed that infinitely small wavelengths would be amplified, generating salt-and pepper patterns for which the
continuum approximation breaks down. Hence, the signs of the loops of Edar and CGTF are inverted, so that the network does not
have a non-diffusible self-activator. Laborious algebraic transformation allowed them to prove that the system can not generate
stationary patterns but that it can undergo oscillatory Turing instabilities. \\
This result is readily recovered using the graph formalism. Stability imposes that $(\sloop{e}\sloopc{red}{b}-\pntwocycle{eb})>0$. 
This means that $a_3(q)$ is formed only by positive terms and therefore cannot fulfill the condition $a_3(q)<0$ . However,
the coefficient $a_2(q)$ can fulfill $a_2(q)<0$, the condition for oscillations. This requires $\sloopc{red}{b}>-\sloop{c}$ and
the following constraint on the diffusion ratio:
\begin{equation*}
a_2(q)= a_2(0)-q^2\diff{c}(\sloopc{red}{b}+\sloop{e})-q^2\diff{b}(\sloop{c}+\sloop{e})+q^4\diff{b}\diff{c}\,<0\,\Rightarrow\,\,\diff{c}/\diff{b}>-\frac{(\sloop{c}+\sloop{e})}{(\sloopc{red}{b}+\sloop{e})}\,>\,1
\end{equation*}
This shows that this network is a oscillatory Turing system of Type-II according to the diffusion constraints.\\
Finally, Klika et al. proposed an alternative network for hair-follicle patterning that is shown in SM fig.\ref{fig:Hair}c. 
The assumption behind this modification is that $CTGF$ does not inhibit the production of $BMP$ directly, but instead inhibits the effects of $BMP$ on $Edar$.
Then they showed that this network can generate stationary patterns. This result is recovered examining $a_3(q)$: 
\begin{equation*}
a_3(q)= a_3(0)+q^2\diff{b}(\sloop{e}\sloop{c}-\pptwocyclec{red}{eb})-q^2\diff{c}\pntwocycle{eb}-\sloopc{black}{e}\cdot q^4\diff{b}\diff{c}<0\,\,\Rightarrow\diff{b}/\diff{c}>\frac{\pntwocycle{eb}}{\sloop{e}\sloop{c}-\pptwocyclec{red}{eb}} >1
\end{equation*}
where the second inequality characteristic of Type-I networks stems from the stability condition $a_2(0)=(\sloop{e}\sloop{c}-\pptwocyclec{red}{eb})-\pntwocycle{eb}>0$.
%The observation that the continuous approximation break-down may imply that the simulations of the %pattering dynamics of Turing filters may not be accurate, particularly for very large wavelengths. Investigation of the computational aspects of this issue is beyond the scope of this work but we have characterized the theoretical foundations of theses systems in SM\ref{Tfilters}.  

\section{Turing filters and Rouche's Theorem}
\label{6th:app} 
%%%%%%%%%%%%%%%%%%%%%%%%%%%%%%%%%%%%%%%%%%%%%%%%%%%%%%%%%%%%%%%%%%%%%%%%%%%%%%%%%%%%%%
This section demonstrates that if all the nodes of the destabilizing module are immobile the network is a Turing filter. 

A Turing filter is a Type-III Turing network characterized by a dispersion relationship
in which the largest eigenvalue converges asymptotically to a positive value for large wavenumbers. 
This means that the network does not generate patterns with a preferred wavelength, because above a certain wavenumber all eigenmodes 
are amplified. Further, the speed of growth of a mode increases with $q$. Because of this, noisy initial conditions evolve to produce 
a noisy pattern of large amplitude and not the typical periodic patterns of a regular Turing network. However, 
initial perturbations with spatial periodicity and a wavenumber above a certain threshold are amplified by these class of networks.
Next we demonstrate that a Turing network with immobile nodes in the destabilizing module and diffusible nodes for the rest is a Turing filter. 
The proof follows the strategy used to demonstrate Theorem 1 in \cite{Cross1978}.\\

\textbf{Theorem.} \textit{A network with a destabilizing module induced by immobile species generates Turing patterns 
independently of the values of the diffusion constants of the rest of the species. 
For} $q\to\infty$ \textit{the largest eigenvalue converges to the largest eigenvalue of the submatrix associated to the  destabilizing module.}\\

\textit{Proof:} Let the destabilizing module be a subgraph of size $K$. The network can be relabeled so that the nodes
that induce it have indexes $\gamma_K=(1,...,K)$. 
The diffusion matrix is then ${\mathbf{D}} = {\text{diag[}}0,..., 0,d_{K+1} ,..,d_N {\text{]}}$, 
where the first $K$ entries correspond to the immobile nodes and $d=(d_{K+1}, ..., d_{N})$ are the diffusion constants of the diffusible species.
In turn, the characteristic polynomial can be written in powers of $q$ as:\\

\begin{equation}
P_q (\lambda ) = \sum\limits_{j = 0}^{N - K - 1} {q^{2j}  \cdot b_j (\lambda ,d ) + q^{2(N - K)} \prod\limits_{i = K + 1}^N {d_i } }  \cdot b_{N - K} (\lambda )
\end{equation}

For $j<N-k$, the coefficients $b_j(\lambda,d_i )$ are polynomials in $\lambda$ that depend on the diffusion constants, whereas
$b_{N - K} (\lambda ) = \det [\lambda {\rm{I - J}}_{\gamma _K }^R ]$  is not a function of the diffusion constants.
From the identity between a matrix determinant and the product of its eigenvalues %$\rm{J_{\gamma _K }^R}$

\begin{equation}
{\rm{det[-J}}_{\gamma _K }^R ] = (-1)^K\prod\limits_{i = 1}^K {\lambda _i }=\sum\limits_{\ell\in I_{\gamma_K}}w(\ell)<0
\end{equation}
where the last inequality uses the fact that the weight of the destabilizing module is negative.
Thus, we conclude that  ${\rm{J}}_{\gamma _K }^R$ has an eigenvalue $\lambda_+$ with positive real part. Therefore, $b_{N - K} (\lambda )$ has 
also zero at $\lambda_+$.
Let $C$ be a closed domain in the right half of the complex plane containing $\lambda_+$. 
For $q$ large enough and independently of $d_i$, the following inequality holds in $C$:
\begin{equation}
\left| {\sum\limits_{j = 0}^{N - K - 1} {q^{2j}  \cdot b_j (\lambda ,d )} } \right| < \left| {q^{2(N - K)} \prod\limits_{i = K + 1}^N {d_i  \cdot b_{N - K} (\lambda )} } \right|
\end{equation}

By Rouche's Theorem, this implies that $P_q(\lambda)$ has also a zero in $C$, that is, an eigenvalue in the right half of the complex plane.
\begin{figure} [!h]
\centering
{\includegraphics[width=.6\textwidth]{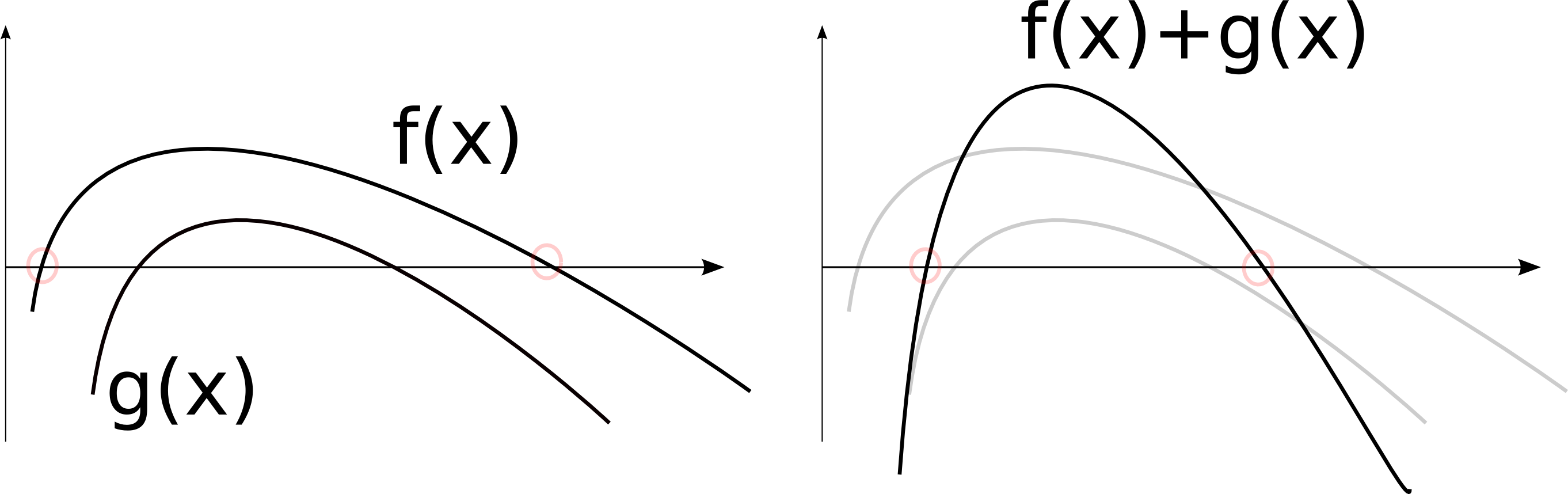}}
\caption{Rouche's theorem states that if $f$ and $g$ are two continuous functions in a closed domain $C$ and $|g(z)| < |f(z)|$ in $C$, then $f$ and $f+g$ have the same number of zeroes in $C$. }
\label{fig:BSW_J13}
\end{figure}
\vspace{0.25cm}
This means that the network will undergo a diffusion-driven instability
for wavenumbers above a certain threshold independently of the values of the diffusion constants. Further, the domain $C$ can be made 
arbitrarily small around $\lambda_+$ and for $q$ large enough the inequality still holds. It follows that an eigenvalue of the full system converges to  
$\lambda_+$ for $q\to\infty$ and concludes the proof.
\begin{figure} [!h] 
\centering
{\includegraphics[width=.9\textwidth]{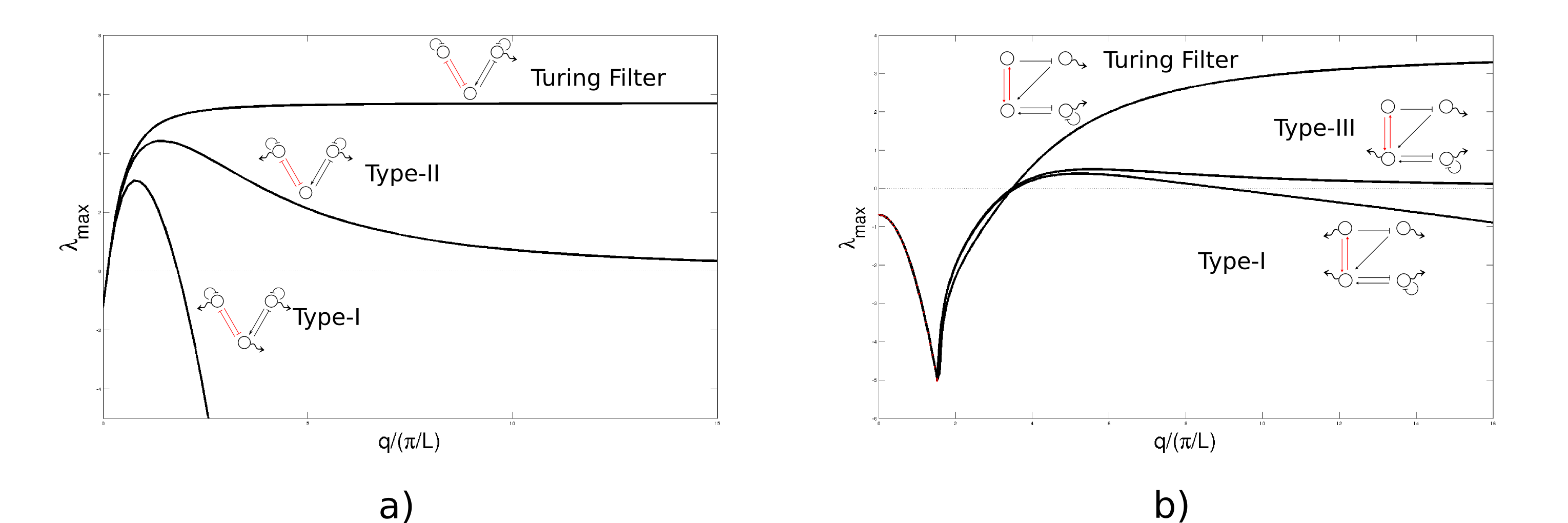}}
\caption{Dispersion relationships of two example Turing networks with all nodes diffusible, one node in
the destabilizing module immobile and all the nodes in the destabilizing module immobile. The latter case produces Turing filters.}
\label{fig:Dispersions}
\end{figure}
\vspace{0.25cm}
To illustrate this result, the variation in the dispersion relationships of two Turing network as more nodes 
of the destabilizing module are assumed to be immobile are shown in SM fig.\ref{fig:Dispersions}.

%%%%%%%%%%%%%%%%%%%%%%%%%%%%%%%%%%%%%%%%%%%%%%%%%%%%%%%%%%%%%%%%%%%%%%%%%%%%%%%%%%%%%%
\section{Properties of Turing pattern phases }
%%%%%%%%%%%%%%%%%%%%%%%%%%%%%%%%%%%%%%%%%%%%%%%%%%%%%%%%%%%%%%%%%%%%%%%%%%%%%%%%%%%%%%

First we will proof that the number of Turing patterns that can be formed with $N$ species is $2^{N-1}$.
This simply requires counting the number of different ways in which $N$ species can be separated in two phases.\\
Thus, we have to count the patterns with $N$ species in one phase and $0$ in the other (i.e, $\binom{N}{0}=1$ ),
with $N-1$ species in one phase and $1$ species in the other (i.e, $\binom{N}{1}=N$ ), 
with $N-2$ species in one phase and $2$ species in the other (i.e, $\binom{N}{2}=N\cdot(N-1)/2$ ) and so forth. 
For a network with an odd number of species, this number is: 
\begin{equation}
\label{odd}
\#_{odd}=\binom{N}{0}+\binom{N}{1}+...+\binom{N}{\frac{(N-1)}{2}}=\sum\limits_{i = 0}^{(N-1)/2} {\binom{N}{i}}
\end{equation}

Using the identity $\binom{N}{k}=\binom{N}{N-k}$ and rearranging equation \ref{odd}, we obtain:

\begin{equation}
\label{oddII}
\begin{split}
\#_{odd}=&\frac{1}{2}\left( \left[\binom{N}{0}+\binom{N}{N}\right]+\left[\binom{N}{1}+\binom{N}{N-1} \right]+...+\left[\binom{N}{\frac{(N-1)}{2}}+\binom{N}{\frac{(N+1)}{2}} \right]  \right)\\
        =&\frac{1}{2}\sum\limits_{i = 0}^{N} {\binom{N}{i}}
\end{split}
\end{equation}

Similarly, for a network with an even number of species, the number of possible patterns is: 
\begin{equation}
\label{even}
\#_{even}=\binom{N}{0}+\binom{N}{1}+...+\binom{N}{\frac{N}{2}-1}+\frac{1}{2}\binom{N}{\frac{N}{2}}=\sum\limits_{i = 0}^{N/2-1} {\binom{N}{i}}+\frac{1}{2}\binom{N}{\frac{N}{2}}
\end{equation}

Rearranging this expression as in the previous case we obtain: 

\begin{equation}
\label{evenII}
\begin{split}
\#_{even}=&\frac{1}{2}\left( \left[\binom{N}{0}+\binom{N}{N}\right]+...+\left[\binom{N}{\frac{N}{2}-1}+\binom{N}{\frac{N}{2}+1} \right]  \right)+\frac{1}{2}\binom{N}{\frac{N}{2}}\\
         =&\frac{1}{2}\sum\limits_{i = 0}^{N} {\binom{N}{i}} 
\end{split}
\end{equation}

The identity $\sum\limits_{i = 0}^N {\binom{N}{i}}=2^N$ follows from the binomial formula. Introducing it in eqs.\ref{oddII}-\ref{evenII} we arrive at 
the final result:

\begin{equation}
\#_{odd}=\#_{even}=\frac{1}{2}\sum\limits_{i = 0}^{N} {\binom{N}{i}}=2^{N-1}
\end{equation}

Next we demonstrate how to construct the networks that generate the patterns with each of the distribution of species in two phases. 
Further, this allows us to show how to construct all the networks that generate each of the $2^{N-1}$ Turing patterns that can be formed by $N$ species.\\

\textbf{Theorem} \textit{The Turing network obtained by inverting the signs of all the incoming and outgoing edges of a particular node generates a 
pattern in which that node switches phase and the rest have the same phase than in the original. This transformation also preserves 1) the speed of pattern appearance  2) the wavelength of the pattern and the amplitude of the maximum and minimums of all the species. }\\

\textit{Proof.} The distribution of the species of a Turing network in two phases is determined by the 
eigenvector associated to the eigenvalue branch of ${\bf{F}^{RD}}(q)={\bf{J}}^R(r_0)-q^2{\bf{D}}$ that has the maximum value for some $q>0$. 
Let $\lambda_q$ and ${\bf{v}}(\lambda_q) = [v_1 ,...,v_i ,...,v_N ]^T$ be an eigenvalue and associated eigenvector of ${\bf{F}}^{{\bf{RD}}}(q)$:

\begin{equation}
\label{eq:eigenProblem}
{\bf{F}}^{{\bf{RD}}}(q)  \cdot {\bf{v}} = \lambda_q   {\bf{v}}
\end{equation}

Precisely, two species are in phase if their components in the eigenvector have the same sign. 
Thus, the proof of this proposition requires showing the relationship  between the eigenvector associated to $\lambda_q$ in 
the original and transformed networks.\\
We define the reflection matrix $L_I$ as the diagonal matrix that has $-1$ in the $i$-th diagonal entry and $+1$ in the rest. 
Note that from the definition of the reflection matrix, $L_I$ is its own inverse:
\begin{equation}
L_I^{ - 1}\cdot L_I  = L_I \cdot L_I^{ - 1}={\bf{I}}
\end{equation}

Applied to a vector ${\bf{v}}$, the reflection matrix $L_I$ inverts the sign of the $i$-th component to ${\bf{v}'}=L_I  \cdot {\bf{v}} = [v_1 ,..., - v_i ,...,v_N ]^T$. Applied to a square matrix ${\bf{A}}$ the transformation ${\bf{A}} \cdot L_I$  inverts the sign of the entries in the $i$-th column. Similarly, the transformation $L_I^{-1}  \cdot {\bf{A}}$  inverts the signs of the entries in the $i$-th row. 
\begin{figure} [!h]
\centering
{\includegraphics[width=0.9\linewidth]{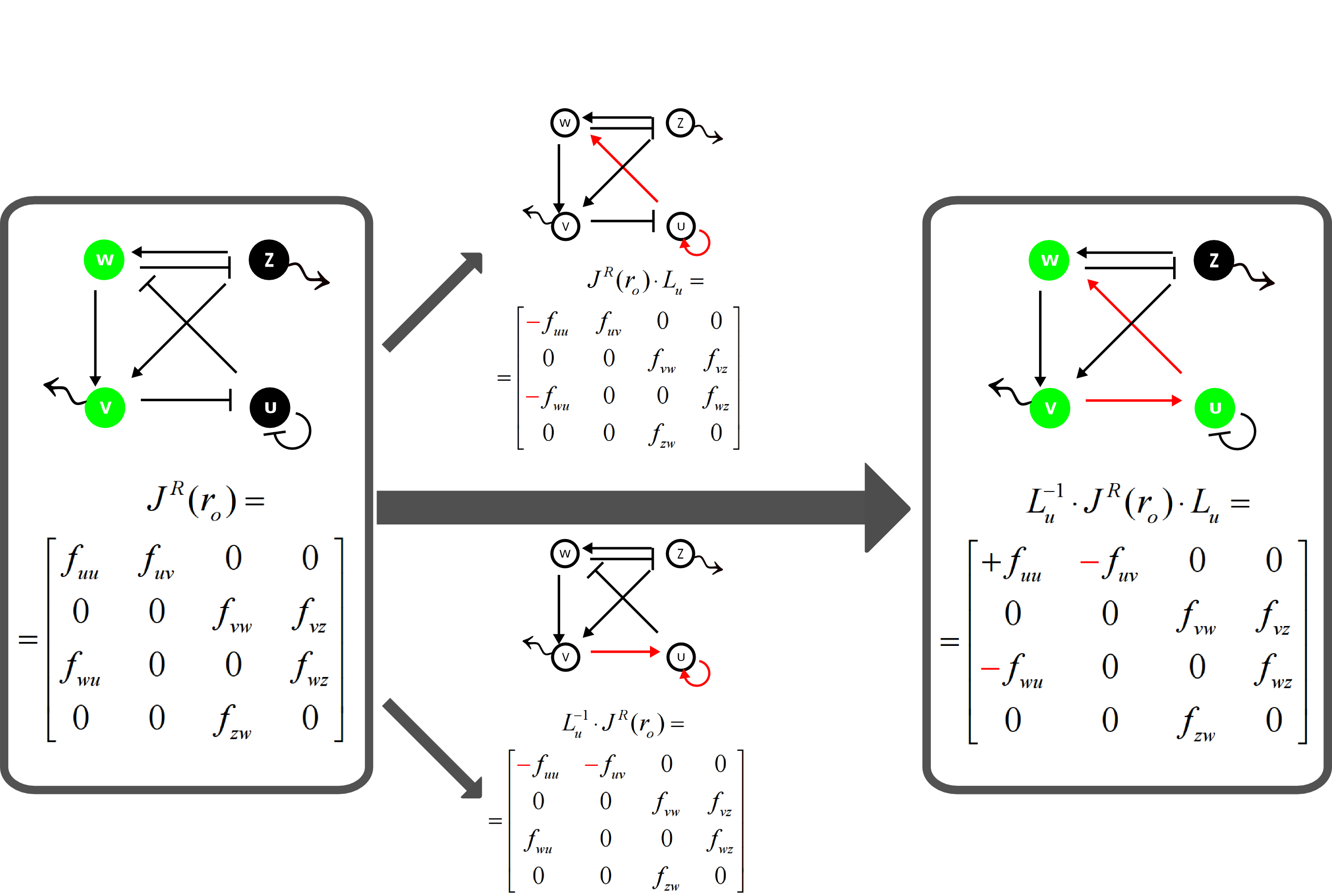}}
\caption{ Illustration of the effect of reflection transformations applied to node $u$ of an example network. Note that only the inversion of both incoming and outgoing edges results in a transformation that preserves the cycles signs. Nodes in phase plotted in the same color.} 
\label{fig:Phases}
\end{figure}
\vspace{0.25cm}
The similarity transformation $L_I^{ - 1}  \cdot {\bf{A}} \cdot L_I$ inverts the signs of the entries along 
the $i$-th column and $i$-th row, leaving the sign of the entry in the diagonal invariant. 
From the definition of the Coates graph of a reaction-diffusion equation, it follows that the transformation ${\bf{F}}^{{\bf{RD'}}} = L_I^{ - 1}  \cdot {\bf{F}}^{{\bf{RD}}}  \cdot L_I$ is equivalent to changing the sign of the all the edges that start or end in the $i$-th node, leaving the loop of node $i$, if it exists, invariant. The relationship between the eigenvectors of ${\bf{F}}^{{\bf{RD}}}$ and ${\bf{F}}^{{\bf{RD'}}}$ can now be obtained from eq.\ref{eq:eigenProblem}:

\begin{equation}
\begin{split} 
{\bf{F}}^{{\bf{RD}}} (q) \cdot ( L_I \cdot L_I^{ - 1})\cdot{\bf{v}} &= \lambda _q  \cdot (L_I\cdot L_I^{ - 1} ) \cdot {\bf{v}}\\
L_I^{ - 1}  \cdot {\bf{F}}^{{\bf{RD}}} (q) \cdot ( L_I \cdot L_I^{ - 1} ) \cdot {\bf{v}}& = \lambda _q\cdot L_I^{ - 1}\cdot (L_I \cdot L_I^{ - 1} )  \cdot {\bf{v}}\\
(L_I^{ - 1}  \cdot {\bf{F}}^{{\bf{RD}}} (q) \cdot  L_I ) \cdot (L_I^{ - 1}  \cdot {\bf{v}}) & = \lambda _q\cdot ( L_I^{ - 1} \cdot {\bf{v}} ) \\
{\bf{F}}^{{\bf{RD'}}} (q) \cdot {\bf{v'}} &= \lambda _q {\bf{v'}}
\end{split}
\end{equation}
Therefore, the eigenvector of the transformed system in which the sign of all the edges that start or end in node $i$ are inverted is:
\begin{equation}
{\bf{v}'}=L_I^{-1}\cdot {\bf{v}}=L_I\cdot {\bf{v}} = [v_1 ,..., - v_i ,...,v_N ]^T
\end{equation} 
This demonstrates that the $i$-the species switches phase and completes the proof. 
The inversion of the signs of the incoming and outgoing edges of a node is a similarity transformation,
which leaves the characteristic polynomial, eigenvalues and dispersion relationship invariant. This means
that the dynamical properties of the original and transformed network are identical because the wavelength,
speed of pattern emergence and amplitudes of the pattern are invariant.  
Further, this type of transformation can be applied repeatedly to any combination of nodes to obtain all 
possible pairwise combinations of species in the two phases. This shows how, from a fixed $N$-node topology and invariant sign distribution of its cycles, 
it is possible to construct all the networks that generate the $2^{N-1}$ possible Turing patterns.\\
As a final remark, is worth emphasizing that this result is general: it does not 
depend on the size of the network and it does not require the network to be minimal. The key element is that
this transformation changes the sign of the edges at node $i$, but it leaves the sign of all the cycles that pass through 
it invariant. This result showcases the power of analyzing Turing networks in this graph-theoretical framework, since it 
provides an almost trivial proof of a result that has not been shown before.

\label{8th:app} 
%%%%%%%%%%%%%%%%%%%%%%%%%%%%%%%%%%%%%%%%%%%%%%%%%%%%%%%%%%%%%%%%%%%%%%%%%%%%%%%%%%%%%%
\section{Rule of thumb to predict pattern phases}
%%%%%%%%%%%%%%%%%%%%%%%%%%%%%%%%%%%%%%%%%%%%%%%%%%%%%%%%%%%%%%%%%%%%%%%%%%%%%%%%%%%%%%
\label{7th:app} 
The topology of a minimal Turing network allows
to predict the phases of the pattern produced. 
The simple method that we have found is based on examining the cycles of the network.
Hence, it does not require to calculate the eigenvectors of the associated eigenvalue problem. 
This particularly advantageous in networks with more than two species, where the analytical
solution of the eigenvalue problem is either very cumbersome or, for $N\geq5$ cannot be found in terms of radicals. 
Numerical solutions can of course always be found, but they do not provide an intuitive understanding. 
In addition, in many real systems the values of the parameters are not known.

    \begin{figure} [!h] 
    \centering
      {\includegraphics[width=0.99\linewidth]{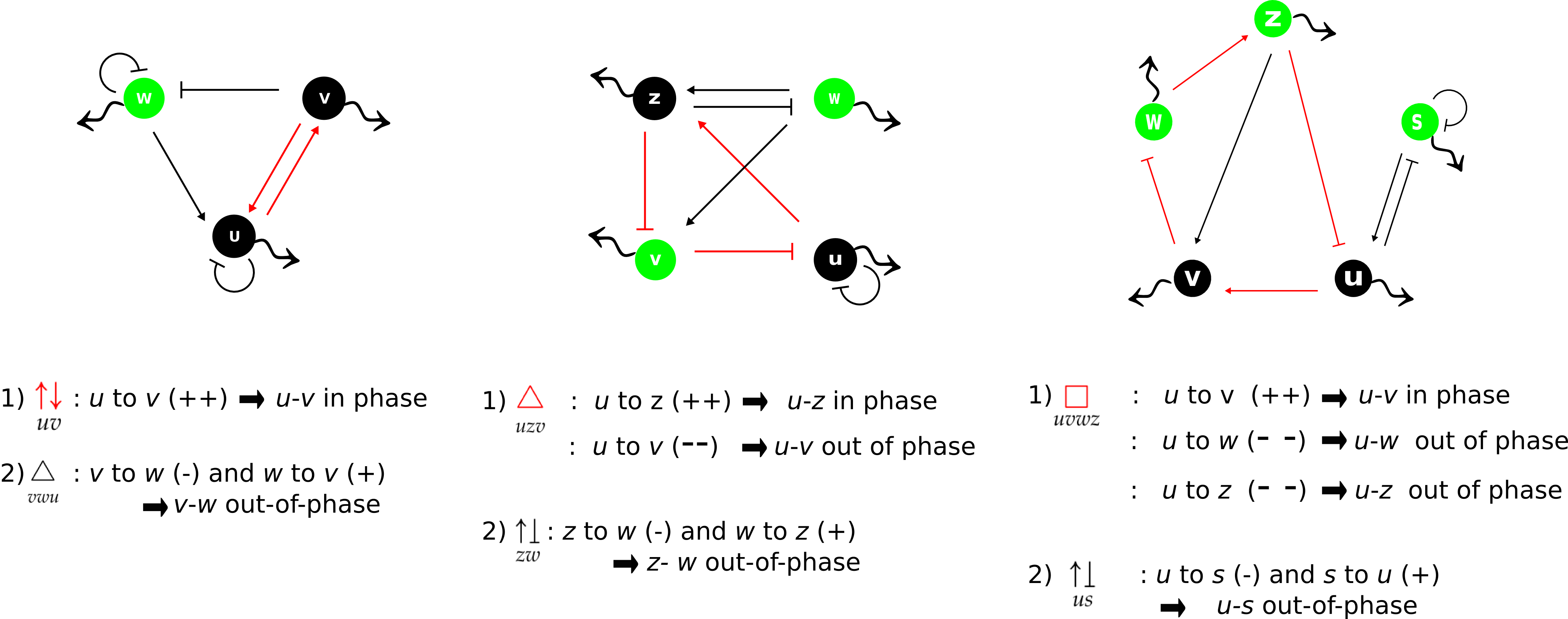}}
      \caption{Examples of the application of the method to establish the phases of the Turing pattern without calculation of the eigenvectors. Positive cycle in destabilizing module in red. Nodes in-phase are filled in the same color. } 
      \label{fig:RuleThumb}
      \end{figure}
      \vspace{0.25cm}

The phase of the species in the positive cycle of the destabilizing factor
is established first. For any two nodes $a$ and $b$, the positive cycle connecting them can be 
divided into two directed paths, one from $a$ to $b$ and another from $b$ to $a$. Also necessarily, 
since the cycle is positive, these paths must either be both positive or both negative. In the former case $a$ and $b$ will be in phase,
in the later, out-of-phase. By this simple procedure, 
the relative phase of every pair of species in the destabilizing module is established.\\  
Next, the phases of the species outside the destabilizing module relative to those inside will be established.
Select a node outside the destabilizing module and find a cycle that 
connects it with a node inside. If the cycle is positive, the phase can be established as before. If the cycle is negative,
again there are just two possibilities for any two nodes $a$ and $b$ in it: either the path from $a$ to $b$ is negative and 
the path from $b$ to $a$ is positive or the opposite. Let $a$ be the species in the destabilizing module and $a$ the specie outside.
If the path from $a$ in to  $b$ is positive, then $b$ is in phase with $a$. If the path from  $a$ to $b$ is 
negative, then $b$ is out-of-phase with $a$. By iteration of this procedure, it is then possible to establish 
the phase of all the remaining nodes in the network.
This method allows to predict correctly the phases of minimal networks of up to $5$ nodes, as shown in SM fig.\ref{fig:RuleThumb}.
In non-minimal networks, there may be more than one cycle of the same order connecting the nodes outside the destabilizing 
module with the nodes that induce it. In this case, there may be an ambiguity if the different connecting cycles have different signs. 
Then, the cycle chosen to establish the phase of nodes outside the destabilizing module should have the largest weight amongst them. 
We have validated this method with only a few examples of non-minimal networks. However, it remains to be proved rigorously in the 
general case of non-minimal networks.

%%%%%%%%%%%%%%%%%%%%%%%%%%%%%%%%%%%%%%%%%%%%%%%%%%%%%%%%%%%%%%%%%%%%%%%%%%%%%%%%%%%%%%
\section{Algebraic proof of the impossibility of Turing patterns in systems with all 
species diffusing at the same rate }
%%%%%%%%%%%%%%%%%%%%%%%%%%%%%%%%%%%%%%%%%%%%%%%%%%%%%%%%%%%%%%%%%%%%%%%%%%%%%%%%%%%%%%
\label{9th:app}
A reaction-diffusion network in which \textit{all} species diffuse requires differential diffusivity. This is a well 
known property of Turing networks;
here we provide a simple proof that carries over for networks of any size. This proof is also meant to highlight 
that the algebraic method does not provide an 
insight into how the diffusion constraints of Turing systems can be relaxed.\\
The emergence of Turing patterns is determined by the sign of the eigenvalues of the following equation 
 
\begin{equation}
P_q(\lambda')=\det [\lambda ' \cdot {\bf{I}} - ({\bf{J}}^R (r_0 ) - q^2 \cdot {\bf{D}})]=0 
\end{equation}

where ${\bf{J}}^R (r_0 )$ is the jacobian of the reaction term in the reaction-diffusion equations.
Let all the species in the network diffuse at the same rate $d$; the diffusion term is the diagonal matrix ${\bf{D}} = d \cdot {\bf{I}}$. Then, 
the eigenvalue problem can be rewritten as:
\begin{equation}
\label{eq:P1}
P_q(\lambda')=\det [\lambda ' \cdot {\bf{I}} - ({\bf{J}}^R (r_0 ) - q^2 d \cdot {\bf{I}})] = \det [(\lambda ' + q^2 d) \cdot {\bf{I}} - {\bf{J}}^R (	r_0 )]=0 
\end{equation}
In turn, the eigenvalues of ${\bf{J}}^R (r_0 )$ are given by:
\begin{equation}
\label{eq:P0}
P_0 (\lambda)=\det [\lambda  \cdot {\bf{I}} - ({\bf{J}}^R (r_0 )]=0
\end{equation}
A Turing network is by definition stable without diffusion. This means that the eigenvalues of ${\bf{J}}^R (r_0 )$  have all negative real part.
Comparing eqs.\ref{eq:P1}-\ref{eq:P0}, it follows that  $\lambda ' = \lambda  - q^2 d < \lambda  < 0$ for $q>0$, which completes the proof.

%%%%%%%%%%%%%%%%%%%%%%%%%%%%%%%%%%%%%%%%%%%%%%%%%%%%%%%%%%%%%%%%%%%%%%%%%%%%%%%%%%%%%%
\section{On Turing minimal networks}
%%%%%%%%%%%%%%%%%%%%%%%%%%%%%%%%%%%%%%%%%%%%%%%%%%%%%%%%%%%%%%%%%%%%%%%%%%%%%%%%%%%%%%
\label{10th:app}
\subsection{Conjecture about number of edges of minimal network }
A minimal Turing network is a strongly connected network that has the minimum number of edges required to 
build a network that can be stable without diffusion and that can undergo diffusion-driven instabilities. 
The strongly connected property is imposed to discard networks with modules that are a readout of the rest 
of the network\footnote{A network is strongly connected if for each pair of nodes $u$ and $v$ there is a path form $u$ to $v$ and vice-versa. 
Equivalently, a network is strongly connected if it cannot be partitioned into two subsets $S$ and $T$ such that all the edges between them
have the initial node in $S$ and the terminal node in $T$ \cite{Brualdi08}. This shows why disregarding networks that are not strongly 
connected allows to filter out networks with modules that are just a downstream readout of a core network}.
In a network with $N$ nodes, stability without diffusion requires that there is at least one stabilizing $\ell_R$-subgraph of every size up to $N$.
Additionally, a network that generates stationary Turing patterns must have a stabilizing $\ell_R$-subgraph of size $K<N$ that forms 
the destabilizing module. Examples of minimal networks obtained with software RDNets \cite{Marcon2016} are shown in the following figure:

    \begin{figure} [!h]
    \centering
      {\includegraphics[width=0.9\linewidth]{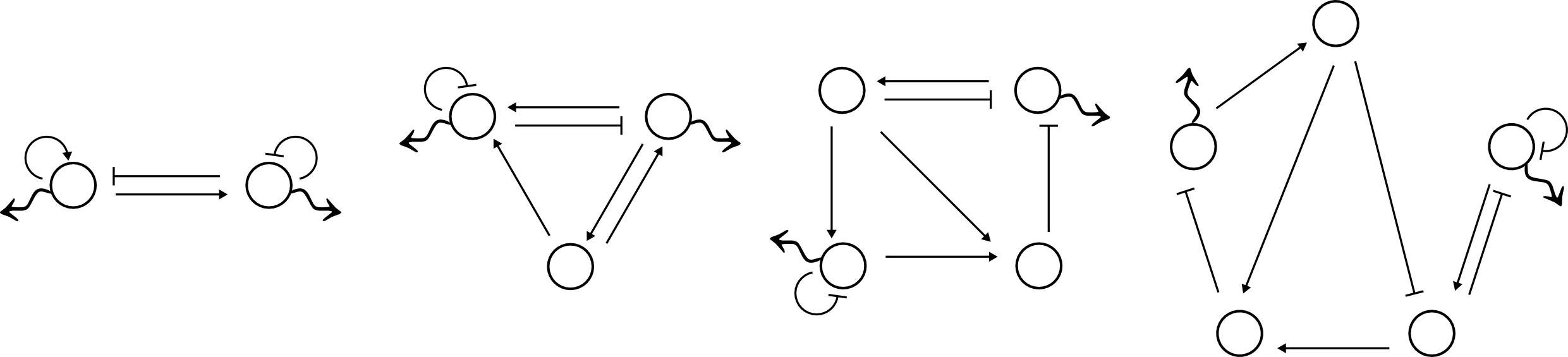}}
      \caption{Examples of minimal Turing networks of $N=2,...,5$ nodes} 
      \label{MinimalTopologies}
      \end{figure}
      \vspace{0.25cm}
Interestingly, we find that in minimal Turing networks, the number of nodes $N$, the number of edges $E$ and
the number of cycles $C$ are given by the following expression:

\begin{equation}
 \label{eq:recurrence}
E=N+C-1
\end{equation}

Thus, beyond the well known case of 2-node networks (which have 4 edges and 3 cycles), we find, surprisingly, 
that to build a minimal Turing network of $N=3,\,4,\,5$ nodes the number of cycles required is 4 in all cases (and the minimal number of edges is $6,\,7$ and $8$ ).  
It is outside the scope of this work to asses whether the recurrence formula holds 
for $N>5$. We leave it then as a conjecture\footnote{Our attempts to build a minimal network of 6 nodes 
with four cycles have been unsuccessful. It is likely that this is not possible, but the relationship between $E, N, C$ might still hold
if 5 cycles and 10 edges form the minimal Turing networks of 6 nodes.}, noting the intriguing parallel with Euler's 
formula for the number of faces as a function of edges and faces in platonic solids and 
its connection with planar graphs \cite{Coxeter1973}.

 \begin{figure} [!h]
 \centering
      {\includegraphics[width=0.99\linewidth]{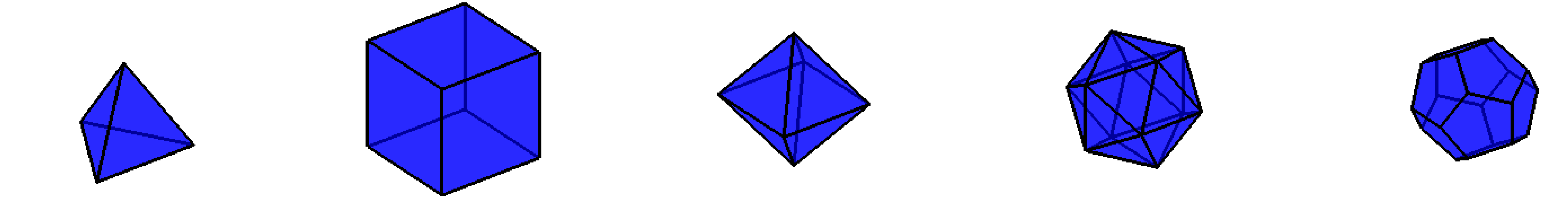}}
      \caption{The number of faces $F$,vertex $V$, and edges $E$ in the 5 platonic solids are given by Euler's formula: $E=F+V-2$. Plot 
      produced using Matlab script by Kevin Moerman.} 
      \label{plato}
      \end{figure}

However, aside from the interest as a theoretical problem, the fact that minimal Turing networks of $N=3,\,4,\,5$
have the same number of cycles should have interesting implications for their robustness. 
Cycles are the true variables that  determine stability and diffusion-driven instabilities rather than individual kinetic parameters. 
However, the number of Hurwitz conditions associated to them grows with $N$. 
It then follows that larger minimal networks have the same number of degrees of freedom to fulfill a larger number of constrains
(or in other words, that the parameters of larger Turing networks should be much more constrained). 
Interestingly, the fact that the stability space is a 4 dimensional space for $N=3,\,4,\,5$  could be used to compare 
the relative robustness of networks of different size.   
%\subsection{Calculation of the volume of the Stability and Turing space of minimal Topologies}
%The calculation of the volume of the Stability space  minimal Topologies takes advantage of the fact that this space is determined by the weight of the 
%network cycles exclusively. For $3 \leq N \geq 5$, all minimal topologies have exactly four cycles. 
%One of them must be the detsbilizing module.
%Hence, we calculate volume of the stability space by varying 
%the absolute value of the weight of each cycle in the range $[0,1]$ and integrating the parameter sets that are compatible with stability. 
%We fix the signs of the cycles, even though for each topology there is at least 
\subsection{Turing topological families of 4 nodes}
    There are 12 different minimal topological families of 4 nodes.
    
      \vspace{0.25cm}

As in the 3-node case, there are important variations between the Stability space of the different families. 
Also like in the 3-node case, setting one or more nodes as non-diffusible allows to relax
the diffusion constraints. It is worth mentioning that this modification does not have the 
same effect in all families. For example, as shown before, only the $T_B$ and $T_J$ families 
have the specific structural properties that allow to obtain a Type-III networks with just one immobile node. 
Generally, while all topological families produce Type-I networks if all nodes are diffusible, the number of networks of other Types, 
Turing Filters and Turing oscillators obtained if nodes are immobile depends on the topological family. Importantly, strong relaxation
of diffusion constrains and large robustness seems to be characteristic of specific families, whereas other 
families have small stability spaces or cannot produce Type-III networks. These features should be taken into account  
when choosing a design for an experimental Turing system.

\bibliographystyle{unsrt}  % {unsrt} {plain}  {apalike}
\bibliography{dFree}